\documentclass[11pt,a4paper]{article}
\pdfoutput = 1
\usepackage{jcappub}
\usepackage[utf8]{inputenc}
\usepackage{amsmath,amssymb}
\usepackage{graphicx}
\usepackage{hyperref}
\usepackage{natbib}

\newcommand{\bq}{\textbf{q}}
\newcommand{\bx}{\textbf{x}}
\newcommand{\bk}{\textbf{k}}
\newcommand{\bPsi}{\boldsymbol{\Psi}}

\newcommand{\vb}[1]{\mathbf{#1}}

\def \dtq{\int d^3 \bq \ }

\newcommand{\avg}[1]{\ensuremath{\left\langle \,#1\, \right\rangle}}

\def\Mpc{\, h^{-1} \, {\rm Mpc}}

\def\kMpc{\, h \, {\rm Mpc}^{-1}}

\def\hq{\hat{q}}

\title{The Reconstructed Power Spectrum in the Zeldovich Approximation}
\author[a]{Shi-Fan Chen}
\author[b]{Zvonimir Vlah}
\author[a,c]{Martin White}
\affiliation[a]{Department of Physics, University of California,
Berkeley, CA 94720}
\affiliation[b]{Theory Department, CERN, 1 Esplanade des Particules, 
CH-1211 Gen\' eve 23, Switzerland}
\affiliation[c]{Department of Astronomy, University of California,
Berkeley, CA 94720}

\emailAdd{shifan\_chen@berkeley.edu}
\emailAdd{zvonimir.vlah@cern.ch}
\emailAdd{mwhite@berkeley.edu}

\date{February 2019}

\abstract{
Density-reconstruction sharpens the baryon acoustic oscillations signal by undoing some of the smoothing incurred by nonlinear structure formation. In this paper we present an analytical model for reconstruction based on the Zeldovich approximation, which for the first time includes a complete set of counterterms and bias terms up to quadratic order and can fit real and redshift-space data pre- and post-reconstruction data in both Fourier and configuration space over a wide range of scales. We compare our model to n-body data at $z = 0$ from the {\tt DarkSky} simulation \cite{skillman14}, finding sub-percent agreement in both real space and in the redshift-space power spectrum monopole out to $k = 0.4 \kMpc$, and out to $k = 0.2 \kMpc$ in the quadrupole, with comparable agreement in configuration space. We compare our model with several popular existing alternatives, updating existing theoretical results for exponential damping in wiggle/no-wiggle splits of the BAO signal and discuss the usually-ignored effect of higher bias contributions on the reconstructed signal. In the appendices, we re-derive the former within our formalism, present exploratory results on higher-order corrections due to nonlinearities inherent to reconstruction, and present numerical techniques with which to calculate the redshift-space power spectrum of biased tracers within the Zeldovich approximation.
}

\begin{document}

\maketitle

\section{Introduction}

Density field reconstruction \cite{ESSS07} is a means of improving the determination of the distance-redshift relation using baryon acoustic oscillations (BAO) \cite{Wei13}.  The BAO method is a ``standard ruler'' test which seeks to measure the scale of a feature in the 2-point function whose physical size is known.  Comparison with the observed size of this feature gives the angular diameter distance and Hubble parameter as a function of redshift.
While the large size of the BAO feature ($100\,$Mpc) makes it relatively immune to systematic effects, nonlinear evolution erases the the oscillations on small scales, or broadens the peak in the correlation function, and reduces the accuracy with which the scale can be measured \cite{Meiksin99,Seo05,White05,Seo07,ESW07}.  However much of the peak broadening comes from motions sourced by very long wavelength fluctuations \cite{ESW07} which are well measured by surveys aiming to measure BAO.  This insight led ref.~\cite{ESSS07} to propose that density-field reconstruction could be applied to regain much of the information lost to non-linearities.
It has been used in all recent BAO surveys to improve their constraints (e.g.~see ref.~\cite{BOSS_DR12} and references therein).

BAO reconstruction has been studied both numerically \cite{Xu12,VM14,VM18} and analytically \cite{ESSS07,PWC09,Noh09,TasZal12,Whi15,Sch15,Seo16,Coh16,Schmittfull17,Hik17,Yu17,Feng18,Modi18,Hada18,Ding18,Sherwin19}.
Our work builds upon these analytic calculations.  Where earlier work made simplifications aimed at highlighting important physical effects, neglected complications such as redshift-space distortions, applied heuristics or otherwise simplified the calculations for explanatory effect, we aim to produce a consistent dynamical theory which can be compared directly to upcoming observational data.  Hence we generalize these calculations to also consider the power spectrum and extend the model to include the complete set of quadratic bias terms.  To our knowledge this is the first dynamical model with a full bias scheme that can produce consistent real and redshift-space results in both Fourier and configuration space, allowing it to be used for consistent fitting of upcoming data.

There has been significant theoretical work on reconstruction since the first algorithm \cite{ESSS07} was suggested.  Most recently, a variety of iterative or alternative reconstruction approaches have been developed \cite{TasZal12,Yu17,Schmittfull17,Feng18,Modi18,Hada18}.  Though our calculations give some insights into these methods, for near-future experiments and for BAO scales these iterative methods do not lead to significant improvements and so we defer consideration of these more complex algorithms to future work.

The outline of this paper is as follows. Section \ref{sec:LPT} reviews the formalism of Lagrangian perturbation theory within which we work.  Section \ref{sec:recon} describes the reconstruction algorithm we seek to model, while Section \ref{sec:results_pk} gives our results in Fourier space, comparing to the configuration-space results where appropriate (Section \ref{sec:config}).  We discuss alternative statistics in Section \ref{sec:omega}.  To assess the range of validity of our models we compare to N-body simulations in Section \ref{sec:nbody}.  A comparison with earlier work is given in Section \ref{sec:earlier} before we conclude in Section \ref{sec:conclusions}.  Some technical details are elaborated in the appendices.

\section{Lagrangian Perturbation theory}
\label{sec:LPT}

The Lagrangian framework \cite{Zel70,Buc89,Mou91,Hiv95,Ber02,Mat08a,Mat08b,CLPT} describes cosmological structure formation by tracking the displacements $\bPsi(\bq)$ of infinitesimal parcels of the matter fluid from their initial (Lagrangian) positions $\bq.$ In this picture the present day matter over- and underdensities are a result of the clustering of the displaced Eulerian positions $\bx(\bq,\tau) = \bq + \bPsi(\bq,\tau)$. The displacements follow the equation of motion $\bPsi''(\bq) + \mathcal{H} \bPsi'(\bq) = - \nabla_x \Phi(\bx),$ where $\Phi(\bx)$ is the gravitational potential which is in turn sourced by the clustered matter fluid via Poisson's equation $\nabla^2 \Phi(\bx,\tau) = \frac{3}{2} \Omega_m(\tau) \mathcal{H}^2(\tau) \delta(\bx,\tau)$ with $\tau$ the conformal time.  This set of equations can be solved perturbatively in terms of the linear overdensity, $\delta_0$, and the first order solution is given by $\bPsi = -D(\tau)\mathbf{\nabla} \nabla^{-2}  \delta_0,$ where $D(\tau)$ is the linear growth factor \cite{Zel70}.

The Lagrangian picture treats tracer bias and advection separately. Given a biased tracer, $a$, with initial overdensity $F^a(\bq) = F^a\left[\partial^2 \Phi(\bq),... \right]$,
the time-evolved tracer overdensity at conformal time $\tau$ is given by number conservation as \cite{Mat08b}
\begin{equation}
    1 + \delta^{a}(\bx,\tau) = \dtq F^{a}(\bq)\ \delta_D\left(\bx - \bq - \bPsi(\bq,\tau)\right)
    \quad .
\end{equation}
The cross power spectrum between two biased tracer populations $a$ and $b$ is then
\begin{equation}
    P^{ab}(k) = \dtq e^{i\bk \cdot \bq}\ \langle F^a(\bq_2) F^b(\bq_1)\ e^{i \bk \cdot \Delta^{ab}} \rangle_{q = |q_2 - q_1|}
    \quad,\quad
    \Delta^{ab} = \bPsi^{a}(\bq_2) - \bPsi^b(\bq_1),
    \label{eqn:cross_spec_1}
\end{equation}
where we have used that the integrated expectation value can only depend on $\bq = \bq_2 - \bq_1$, due to the translation invariance of the underlying theory.
The bias functionals, $F^{a,b}$, can be Taylor expanded in terms of bias coefficients
\begin{equation}
    F^{a}(\bq) = 1 + b^a_1 \delta_0(\bq) + \frac{1}{2} b^a_2 \left(\delta_0(\bq)^2 - \langle \delta_0^2 \rangle\right) + b_s^a \left( s^2(\bq) - \langle s^2 \rangle\right) + b^a_{\nabla^2} \nabla^2_q \delta_0(\bq) + \cdots ,
\end{equation}
where $s^2 = s_{ij} s_{ij}$ is the square of the shear field, i.e.\ the traceless part of $\partial\partial\Phi$. Following ref.~\cite{VlaCasWhi16}, we also consider contributions from a ``derivative bias'' $b_{\nabla^2}$, i.e.\ corrections to the bias expansion at scales close to the halo radius $R_h$ proportional to $\nabla^2 \delta_0$; such contributions will, however, be essentially degenerate with counterterms renormalizing nonlinearities in the Zeldovich power spectrum and we will therefore not enumerate them separately in the rest of this work unless otherwise stated.

In this work our focus will be on modelling reconstruction within the Zeldovich approximation \cite{Zel70,Whi14}, which keeps only the linear order term in the dynamics of $\bPsi$ but re-sums the effects of the displacement to all orders in a Galilean-invariant manner (this is true for reconstruction also if we take it to mean that all displacements transform the same way). This is specifically accomplished by evaluating the exponential in Equation~\ref{eqn:cross_spec_1} via the cumulant expansion, and evaluating the bias expansion using functional derivatives (see e.g.\ refs.~\cite{Mat08b,CLPT,VlaCasWhi16}).  Following standard techniques, as outlined in the references above, the resulting expression for the cross spectrum is
\begin{align}
    P^{ab}(k) =  &\dtq e^{i\bk \cdot \bq}\ e^{-\frac{1}{2}k_i k_j A^{ab}_{ij}}\
    \Big[ 1 + \alpha_0 k^2 + i b_1^b \bk \cdot U^{a} + i b_1^a \bk \cdot U^{a} + b_1^a b_1^b \xi_L + \frac{1}{2} b_2^a b_2^b \xi_L^2 \nonumber \\
    &- \frac{1}{2}k_i k_j \ (b_2^b U^{a}_i U^{a}_j + b_2^a U^{b}_i U^{b}_j + 2 b_1^a b_1^b U^{a}_i U^{b}_j) + i k_i (b_2^b b_1^a U^{a}_i + b_1^b b_2^a U^{b}_i)\ \xi_L \nonumber\\
    &- \frac{1}{2} k_i k_j (b_s^a \Upsilon^{b}_{ij} + b_s^a \Upsilon^{b}_{ij})   + i k_i (b_1^a b_s^b V^{ab}_i + b_1^b b_s^a V^{ba}_i) \nonumber \\
    &+ \frac{1}{2}(b_2^a b_s^b + b_2^b b_s^a) \chi^{12} + b_s^a b_s^b \zeta + \cdots \Big] 
    \label{eqn:cross_spectra}
\end{align}
where we have defined\footnote{These are generalizations of the similar auto-spectrum quantities defined in refs.~\cite{CLPT,Whi14,VlaCasWhi16}.} the quadratic two point functions
\begin{equation}
       A^{ab}_{ij} = \langle \Delta^{ab}_i \Delta^{ab}_j \rangle, \; U^{b}_i = \langle \Delta^{ab}_i \delta_0(\bq_2) \rangle, \; \xi_L = \langle \delta_0(\bq_2) \delta_0(\bq_1) \rangle
\end{equation}
and shear correlators
\begin{equation}
    \zeta = \langle s^2(\bq_2) s^2(\bq_1) \rangle, \; \Upsilon^{b}_{ij} = \langle \Delta^{ab}_i \Delta^{ab}_j s^2(\bq_2) \rangle, \; V^{ab}_i = \langle \Delta^{ab}_i \delta_0(\bq_2) s^2(\bq_1) \rangle, \; \chi^{12} = \langle \delta^2_0(\bq_1) s^2(\bq_2) \rangle.
\end{equation}
Note that in the above calculations we have, without loss of generality, associated tracers $a$ and $b$ with Lagrangian positions $\bq_2$ and $\bq_1$, respectively.  The quantities in Equation~\ref{eqn:cross_spectra} with $a$ and $b$ swapped can also be calculated by swapping the positions $\bq_1 \leftrightarrow \bq_2$. As an example, $U^b = -\langle \bPsi^b(\bq_1) \delta_0(\bq_2) \rangle$ is the two-point function between the displacement of tracer $b$ and the matter overdensity. The vector and tensor two point functions defined above can be decomposed via rotational symmetry into scalar components, e.g.\ $A_{ij} = X(q) \delta_{ij} + Y(q) \hq_i \hq_j$ and $U_i = U(q) \hq_i$. Formulae for these functions, expressed as Hankel transforms of power spectra, are given in Appendix \ref{sec:correlators}. Finally, we include the contribution $\alpha_0 k^2$ in the square brackets of Equation~\ref{eqn:cross_spectra} as the lowest-order counterterm renormalizing sensitivities to small-scale power in $A_{ij}$ --- in practice this simply modifies the matter contribution $P_{\rm Zel}(k)$ ($\propto 1$ in the square brackets) to $(1 + \alpha_0 k^2) P_{\rm Zel}(k)$ (see e.g.\ refs.~\cite{VlaWhiAvi15,PorSenZal14}). Each term in Equation~\ref{eqn:cross_spectra} can be evaluated as Hankel transforms (see e.g.\ ref.~\cite{VlaCasWhi16}) using the identities given at the end of \cite{VlaWhiAvi15}, which we carry out using the {\tt mcfit} package\footnote{https://github.com/eelregit/mcfit}.

The Lagrangian formalism allows a straightforward translation between real and redshift space via a mapping of the Lagrangian displacements. In particular, assuming the plane-parallel approximation\footnote{This should be an excellent approximation on BAO scales \cite{CasWhi18a}, but if necessary the formalism can be modified to include ``wide-angle'' effects \cite{CasWhi18b}.} and working in the Zeldovich approximation, quantities in redshift space are given simply by substituting $\bPsi_i \rightarrow \bPsi^R_i =  R_{ij} \bPsi_j$ \cite{Mat08b}. Here $R_{ij} = \delta_{ij} + f \hat{n}_i \hat{n}_j$, where $\hat{n}$ denotes the line-of-sight direction and $f = d\ln D/d\ln a$ is the linear-theory growth rate. To lowest order, transforming into redshift space requires the inclusion of a second counterterm dependent on the line-of-sight angle $\nu = \hat{k} \cdot \hat{n}.$ We can see this explicitly, for example, in the UV-sensitive zero-lag term in $A_{ij}$, which gains an angular dependence
\begin{equation}
    k_i k_j \langle (\bPsi_i + \hat{n}_i\hat{n}_l \dot{\bPsi}_l )(\bPsi_j + \hat{n}_j\hat{n}_m \dot{\bPsi}_m ) \rangle \equiv k^2 ( X(0) + (2 \dot{X}(0) + \ddot{X}(0)) \nu^2),
    \label{eqn:zerolag_rsd}
\end{equation}
where $\dot{\bPsi}$ is the velocity in Hubble units equal to $f \bPsi$ in the Zeldovich approximation\footnote{See refs.~\cite{WanReiWhi14, VlaCasWhi16} for a more detailed exposition of the ``dot notation.''}; roughly speaking, we need one angle-independent counterterm $\alpha_0 k^2$ to absorb the UV dependence of $X(0)$ and another $\alpha_2 k^2 \nu^2$ to absorb the UV dependence of the velocities. The complete set of counterterms in redshift space thus makes a contribution of the form $(\alpha_0 + \alpha_2 \nu^2) k^2 P_{ZA}(\bk);$ since $P_{ZA}(\bk)$ is equal to $(1 + f\nu^2)^2 P_L(k)$ to linear order, an equivalent viewpoint---which we will adopt in this work--- is to have constant counterterms $\bar{\alpha}_0 k^2$ and $\bar{\alpha}_2 k^2$ for the monopole and quadrupole, respectively, where the barred counterterms are linear combinations of the unbarred quantities.

\section{Reconstruction algorithm}
\label{sec:recon}

In this section we describe two possible methods for reconstruction in redshift space, both built around the Zeldovich approximation. The standard procedure for reconstruction was developed in ref.~\cite{ESSS07} and involves displacing both observed galaxies and a spatially uniform distribution by a calculated shift field, $\chi$, then taking the relative density contrast between the two sets of particles as the reconstructed density field. For a suitably chosen $\chi$, this can reduce the effect of large scale (IR) bulk flows that ``blur'' the BAO feature. However there is no consensus in the community on the correct procedure for handling redshift-space distortions: the implementation in ref.~\cite{Pad12} chose to multiply $\chi$ by $1+f$ in the line-of-sight direction for $\delta_d$ but not for $\delta_s$.  This `undoes' the supercluster infall effect \cite{Kai87} and reduces the $\ell>0$ moments of the 2-point function on large scales.  Ref.~\cite{Whi15} suggested a symmetric treatment of $\delta_d$ and $\delta_s$, which recovers linear theory on large scales.  This is more natural from the point of view of perturbation theory and better behaved near the boundaries, but is less often implemented on data.  A number of other choices were explored in ref.~\cite{Seo16} but in this work we will restrict our attention to the two methods described above.

The reconstruction procedure consists of the following steps \cite{ESSS07}:
\begin{enumerate}
\item Smooth the observed galaxy density field $\delta_g$ with a kernel $\mathcal{S}$ to filter out small scale (high $k$) modes, which are difficult to model.  We use a Gaussian smoothing of scale $R_s$, specifically $\mathcal{S}(k) =  \exp[-(kR_s)^2/2]$, though none of our analytic results will depend specifically on this choice.  For galaxy surveys Gaussian smoothing has been universally adopted (though with different conventions for $R_s$) but in other contexts it may be advantageous to implement a Wiener filter instead (e.g.~ref.~\cite{Coh16}).
\item Compute the shift, $\mathbf{\chi}$, by dividing the smoothed galaxy density field by a  bias factor $b$ and linear RSD factor \cite{Kai87} and then take the inverse gradient. Assuming linear theory with scale-independent bias and supercluster infall holds on large scales, the calculated shift field should approximate the negative smoothed Zeldovich displacement.
In a simulation with a periodic box, these first two steps can be implemented
using FFTs as
\begin{equation}
  \mathbf{\chi}_{\mathbf{k}} = -\frac{i\mathbf{k}}{k^2}
  \mathcal{S}(k)\ \Big( \frac{\delta_g(\mathbf{k})}{b + f\nu^2} \Big) \approx - \mathcal{S}(k) \bPsi^{(1)}(\bk)
\label{eqn:recon_shift}
\end{equation}
where the bias factor is related to the Lagrangian first-order bias by $b=1+b_1$ and we have defined the line-of-sight angle $\nu = \hat{n} \cdot \hat{k}$.
For non-periodic data the relevant differential equation can be solved by
 multigrid\footnote{https://github.com/martinjameswhite/recon\_code} or by
 linear algebra techniques \cite{Pad12} or iteratively using FFTs \cite{Burden15}.
\item Move the galaxies by $\chi_d = \mathbf{R} \chi$ and compute the ``displaced''
density field, $\delta_d$.
\item Shift an initially spatially uniform distribution of particles by
\begin{itemize}
    \item \textbf{Rec-Sym}: $\chi_s = \mathbf{R}\chi,$ i.e.\ the same amount as the observed galaxies, or,
    \item \textbf{Rec-Iso}: The un-redshifted $\chi_s = \chi$.
\end{itemize}
to form the ``shifted'' density field, $\delta_s$. Note that we have borrowed the nomenclature of ref.~\cite{Seo16} for the latter, which ``isotropizes'' the reconstructed field on large scales.  For the former we use ``Rec-Sym'' to indicate the symmetry of the treatment of $\delta_d$ and $\delta_s$.
\item The reconstructed density field is defined as $\delta_r\equiv \delta_d-\delta_s$ with power spectrum $P_r(k)\propto \langle \left| \delta_r^2\right|\rangle$.
\end{enumerate}
Throughout we shall assume that the fiducial cosmology and halo bias are properly known during reconstruction (see e.g.~refs.~\cite{Sherwin19,Carter19} for relaxation of this assumption), and take the approximation in Eq.~\ref{eqn:recon_shift} to be exact. For further discussion of this point see refs.~\cite{Sch15,Hik17}.
The procedure in real space can be straightforwardly obtained by setting $f = 0$, in which case Rec-Sym and Rec-Iso become equivalent.  Taking the limit $\mathcal{S}\to 0$ or $\chi\to 0$ returns the `raw' spectrum, before reconstruction.

\section{Reconstructed power spectrum}
\label{sec:results_pk}

There has been significant earlier work on modeling density-field reconstruction within perturbation theory \cite{ESSS07,PWC09,Noh09,TasZal12,Whi15,Sch15,Seo16,Coh16,Schmittfull17,Hik17,Yu17,Feng18,Modi18,Hada18,Ding18,Sherwin19}.  In particular ref.~\cite{Whi15} presented a calculation of the configuration-space two-point function (the correlation function) under the assumption of Zeldovich dynamics and that $\chi= -\mathcal{S}\bPsi$.  In this paper we generalize that calculation to a more complete bias model (see \S\ref{sec:config}), including all terms allowed by symmetries up to quadratic order as well as a proper set of counterterms, and we show how to implement the model in Fourier space.  We have explicitly checked that the Hankel transform of our Fourier-space expressions matches the direct configuration-space calculation to 1\% in all terms, and we release code which makes consistent predictions for both statistics with a common set of parameters.  To our knowledge this is the first calculation which provides self-consistent predictions in both spaces, uses a dynamical rather than a heuristic model, works in redshift space and has a full set of bias and counterterms.

Our focus in this section will be to model the reconstructed power spectrum using Lagrangian perturbation theory in both real and redshift space (the expression for the `propagator' is given in Appendix \ref{app:zel_prop} for completeness). Following the algorithm outlined above, the reconstructed power spectrum in real space is given by $P_{\rm recon} = P^{dd} + P^{ss} - 2 P^{ds}.$ Within the Lagrangian framework we can write the displaced density field as
\begin{align}
    1 + \delta_d(\vb{r}) &= \int d^3\bx \ (1 + \delta(\bx))\ \delta_{D}\left[\vb{r} - \bx - \chi_d(\bx)\right] \nonumber \\
    &= \int d^3\bx \dtq F(\bq)\ \delta_{D}\left[\bx - \bq - \bPsi(\bq)\right]\ \delta_{D}\left[\vb{r} - \bx - \chi_d(\bx)\right] \nonumber \\
    &= \dtq F(\bq)\ \delta_D\left[\textbf{r} - \bq - \bPsi(\bq) - \chi_d(\bq + \bPsi(\bq))\right]  ,
\label{eq:rec_map}
\end{align}
where we performed the $\bx$ integral using the first $\delta$-function to go from the second to third lines. Importantly while the fluid displacement, $\bPsi$, is evaluated at the Lagrangian position, $\bq$, the shift field is evaluated at the shifted Eulerian position, $\bq + \bPsi$. The above equalities hold both when the pre-reconstruction coordinate, $\bx$, is in real or redshift space, with the implicit substitution of $\bPsi \rightarrow \mathbf{R}\bPsi$ in the latter case, as long as the appropriate shift field $\chi_d$ is chosen. The expression for the shifted density can be similarly derived or found by setting $\bPsi(\bq) = 0$ and $\chi_d \rightarrow \chi_s$ in the above expression. In Fourier space this translates to
\begin{align}
    (2\pi)^3 \delta_D(\bk) + \delta_d(\bk) &= \dtq e^{-i\bk \cdot \bq}\ F(\bq)\ e^{-i\bk \cdot \big[ \bPsi(\bq) + \chi_d(\bq + \bPsi(\bq)) \big]} \nonumber\\
    (2\pi)^3 \delta_D(\bk) + \delta_s(\bk) &= \dtq e^{-i\bk \cdot \bq}\ e^{-i\bk \cdot \chi_s(\bq)}.
    \label{eqn:deltak_ds}
\end{align}
Below we will make the approximation $\chi(\bq+\bPsi) \approx \chi(\bq).$ The nonlinearities from the Lagrangian-to-Eulerian mapping can be understood as a perturbation series in $\bPsi/R$, where $R$ is the smoothing scale, and we explore their consequences in Appendix \ref{app:EulerianPosition} (see also refs.~\cite{Sch15,Hik17} and the discussion in ref.~\cite{Whi15}). Within this approximation we can treat the displaced and shifted field as tracers with displacements
\begin{equation}
    \bPsi^{d} = \bPsi + \chi_d, \quad \bPsi^s = \chi_s,
    \label{eqn:recon_disps}
\end{equation}
where the Zeldovich displacements should be understood as being in redshift space for the displaced field and in either redshift or real space for the shifted field depending on the method used.  In this picture the ``displaced'' tracer has the same bias functional as the original galaxies ($F^d \equiv F^g$) while the ``shifted'' tracer is unbiased ($F^s\equiv 1$).  A straightforward consequence of the reconstruction procedure is that, like that of any discrete tracer, the shift field autospectrum will contain an independent shot noise term $P^{ss}_{\rm SN} = 1/n_{s},$ where $n_s$ is the number density of the uniform random particles. The full shot noise contribution to the reconstructed spectrum is the the sum of the galaxy and random particle shot noises.

\subsection{Real space}
\label{sec:pk_real}

In real space both the displaced and shifted fields are moved by the same, smoothed negative Zeldovich displacement, $\chi_d = \chi_s = - S \star \bPsi$, such that in Fourier space
\begin{equation}
    \bPsi^d(\bk) = \left[1 - \mathcal{S}(k)\right] \bPsi(\bk), \quad \bPsi^s(\bk) = - \mathcal{S}(k) \bPsi(\bk).
\end{equation}
and the auto- and cross-spectra can be calculated using Equation~\ref{eqn:cross_spectra} and the correlators in Appendix~\ref{sec:correlators}, using linear theory spectra
\begin{equation}
    P^{dd}_L(k) = \left[1 - \mathcal{S}(k)\right]^2 P_L(k), \quad P^{ds}_L(k) = -\mathcal{S}(k) \left[1 - \mathcal{S}(k)\right] P_L(k), \quad P^{ss}_L(k) = \mathcal{S}(k)^2 P_L(k)
\end{equation}
as well as tracer-matter power spectra
\begin{equation}
    P^{dm}_L(k) = \left[1 - \mathcal{S}(k)\right] P_L(k) \quad , \quad
    P^{sm}_L(k) = -\mathcal{S}(k) P_L(k).
    \label{eqn:linspec}
\end{equation}
Note that the shifted field is negatively correlated with both the matter and displaced fields by-construction, since the random particles are displaced in the opposite direction of the (smoothed) Zeldovich displacement.

\begin{figure}
    \centering
    \includegraphics[width=\textwidth]{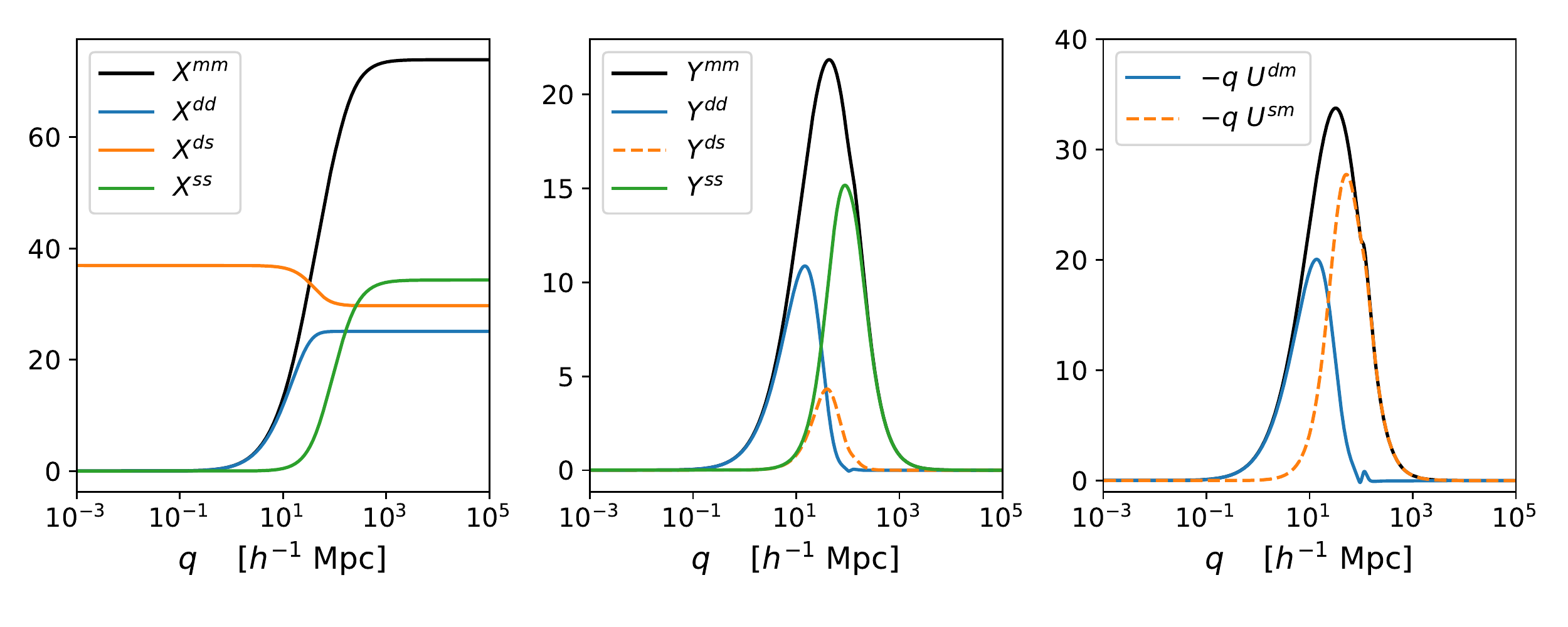}
    \caption{Lagrangian space two point functions used to compute reconstructed power spectra. Dashed quantities have been multiplied by an overall negative sign, and reflect that the shifted field is defined to be negatively correlated with the underlying matter field. Roughly speaking, the shifted and displaced correlators reproduce the general trend for the total matter correlators, shown in black, on large and small scales, respectively. An exception is $X^{ds}$, whose non-vanishing value on small scales reflect that the point values of $\bPsi^d$ and $\bPsi^s$ differ exactly by the Zeldovich displacement. Note also the small but visible features around $q = 100 \Mpc$, i.e. the BAO scale.
    }
    \label{fig:qfuncs_xyu}
\end{figure}

The Lagrangian space two-point correlation functions required to calculate the pre- and post-reconstruction power spectra, normalized to their present-day values, are shown in Figure~\ref{fig:qfuncs_xyu}. For simplicity we have excluded the shear correlators and refer readers to Appendix~\ref{sec:correlators} for further details. The components $X$ and $Y$ describe correlation functions of two displacements, while the $U$'s involve those with only one displacement, such that the former are Hankel transforms of the linear tracer-tracer spectra, while the latter involve the linear tracer-matter spectra. As expected, the $Y$'s and $U$'s for the displaced and shifted fields contain the behavior of the full matter contribution and small and large scales, respectively, and cross correlations between the shifted field and the displaced or matter fields is negative. 

The $X(q)$ components however, especially the cross-correlation $X^{ds}$,  display more subtle behavior. In particular, we have
\begin{equation}
    A^{ds}_{ij}(\bq) \stackrel{q \rightarrow 0}{=} \langle \bPsi^d_i \bPsi^d_j \rangle +   \langle \bPsi^s_i \bPsi^s_j \rangle -  2\langle \bPsi^d_i \bPsi^s_j \rangle \equiv \Sigma^2 \delta_{ij},
    \label{eqn:sigma_def}
\end{equation}
such that $X^{ds}(q) \rightarrow \Sigma^2$ as $q\rightarrow 0$. This is because, when evaluated at the same point, $\bPsi^d - \bPsi^s = \bPsi,$ i.e.\ the difference between the displaced the shifted displacements is none other than the original Zeldovich displacement. This in turn implies that the cross spectrum is damped at small scales $\propto\exp[-k^2\Sigma^2/2]$ due to the nonzero displacement between the displaced and shifted fields. Similar behavior is seen in the evaluation of unequal-time correlation functions \cite{Chisari19} and the baryon-cold dark matter cross-correlation \cite{Lewandowski15,ChenCasWhi19}, though the physical mechanisms are of course different. At large scales, we similarly have
\begin{equation}
    A^{ds}_{ij}(\bq) \stackrel{q \rightarrow \infty}{=} \langle \bPsi^d_i \bPsi^d_j \rangle +   \langle \bPsi^s_i \bPsi^s_j \rangle \equiv \Big( \Sigma_{dd}^2 + \Sigma_{ss}^2 \Big) \delta_{ij},
\end{equation}
such that $X^{ds}$ asymptotes to the average of $X^{dd}$ and $X^{ss}$ at large separations. For completeness, we give explicit expressions for the displaced and shifted $X^{ab}$ here:
\begin{align}
    X^{dd}(q) &= \frac{2}{3} \int \frac{dk}{2\pi^2} \Big[ \  1 - \Big( j_0(kq) + j_2(kq) \Big) \Big]\ \Big(1 - \mathcal{S}(k) \Big)^2 P_L(k) \nonumber \\
    X^{ds}(q) &= \frac{2}{3} \int \frac{dk}{2\pi^2} \Big[ \  \frac{1}{2} \Big( (1 - \mathcal{S}(k))^2 + \mathcal{S}(k)^2   \Big) + \mathcal{S}(k) \Big(1 - \mathcal{S}(k) \Big)\Big( j_0(kq) + j_2(kq) \Big) \Big]\  P_L(k) \nonumber \\
    X^{ss}(q) &= \frac{2}{3} \int \frac{dk}{2\pi^2} \Big[ \  1 - \Big( j_0(kq) + j_2(kq) \Big) \Big]\ \mathcal{S}^2(k) P_L(k).
    \label{eqn:Xds}
\end{align}
The corresponding expressions for $Y^{ab}$ can be directly obtained by calculating $-3$ times the $j_2$ components. As we shall discuss further in Section \ref{sec:earlier}, the signs for the Bessel function coefficients in our expression for $X^{ds}$ differ from those in ref.~\cite{Ding18}.  We note also, as has been emphasized before \cite{PWC09}, each of the three contributions to $P_{\rm recon}$ has a different damping factor which can only be roughly approximated by a single Gaussian term.

\begin{figure}
    \centering
    \includegraphics[width=\textwidth]{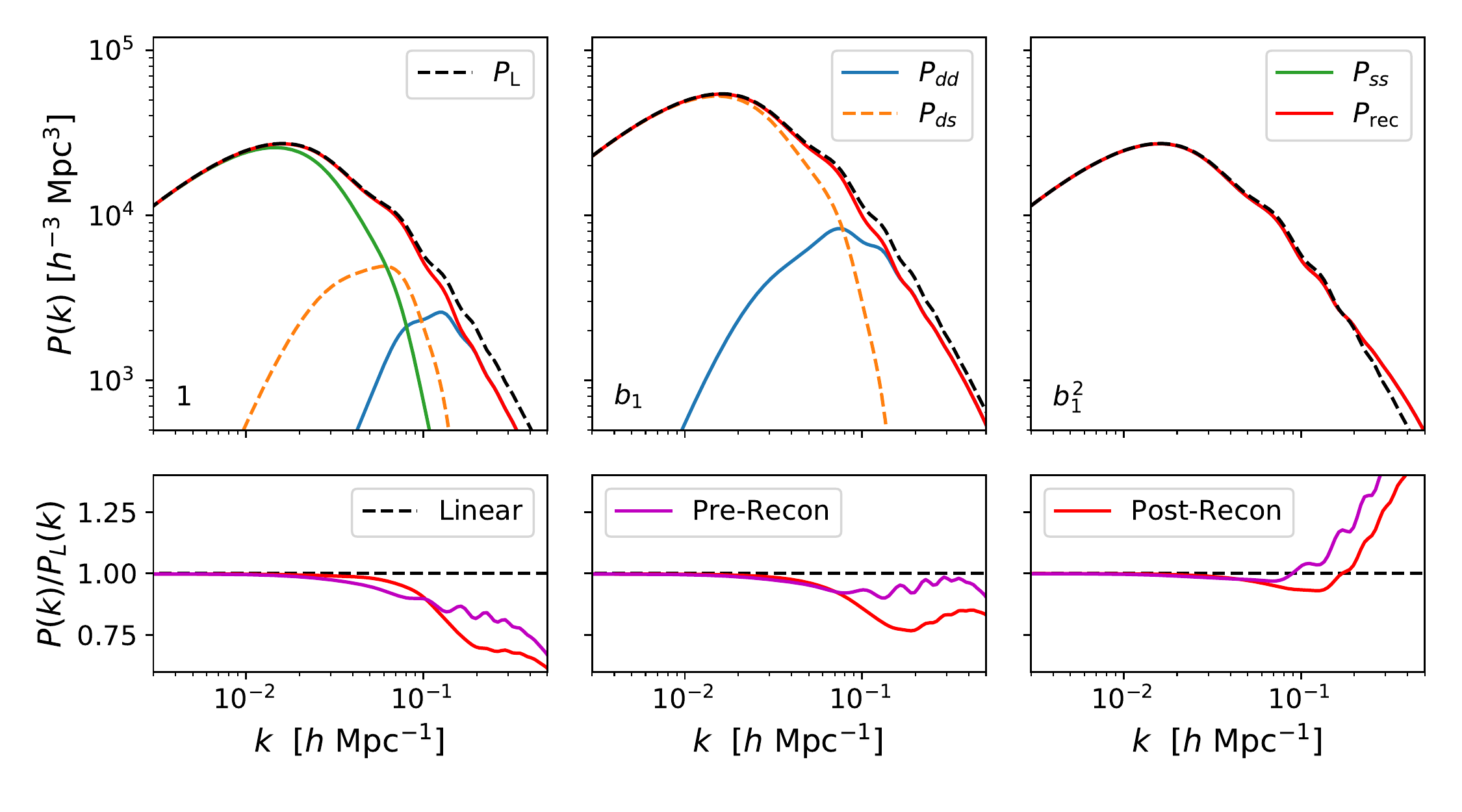}
    \caption{(Top) Real-space power spectra contributions, displaced-displaced, displaced-shifted and shifted-shifted, for the lowest order bias terms $1$, $b_1$, $b_1^2$, and their sum, compared to linear theory at $z = 0$. The pure matter piece is the only term that receives contributions from all three combinations of $d$ and $s$, and the $b_1^2$ term consists only of the $dd$ contribution. All three bias terms tend to linear theory on large scales but exhibit somewhat different broadband behavior at high $k$. (Bottom) The ratio of the above bias terms with the linear theory power spectrum, compared with the pre-reconstruction Zeldovich power spectrum. While both the pre- and post-reconstruction Zeldovich spectra differ with the linear spectrum in the broadband at small scales, the Zeldovich approximation predicts that the the oscillatory features in the reconstructed spectrum are almost identical to those in the linear spectrum, such that the wiggles are almost completely normalized out for the reconstructed spectrum.}
    \label{fig:precon_z0}
\end{figure}

\begin{figure}
    \centering
    \includegraphics[width=\textwidth]{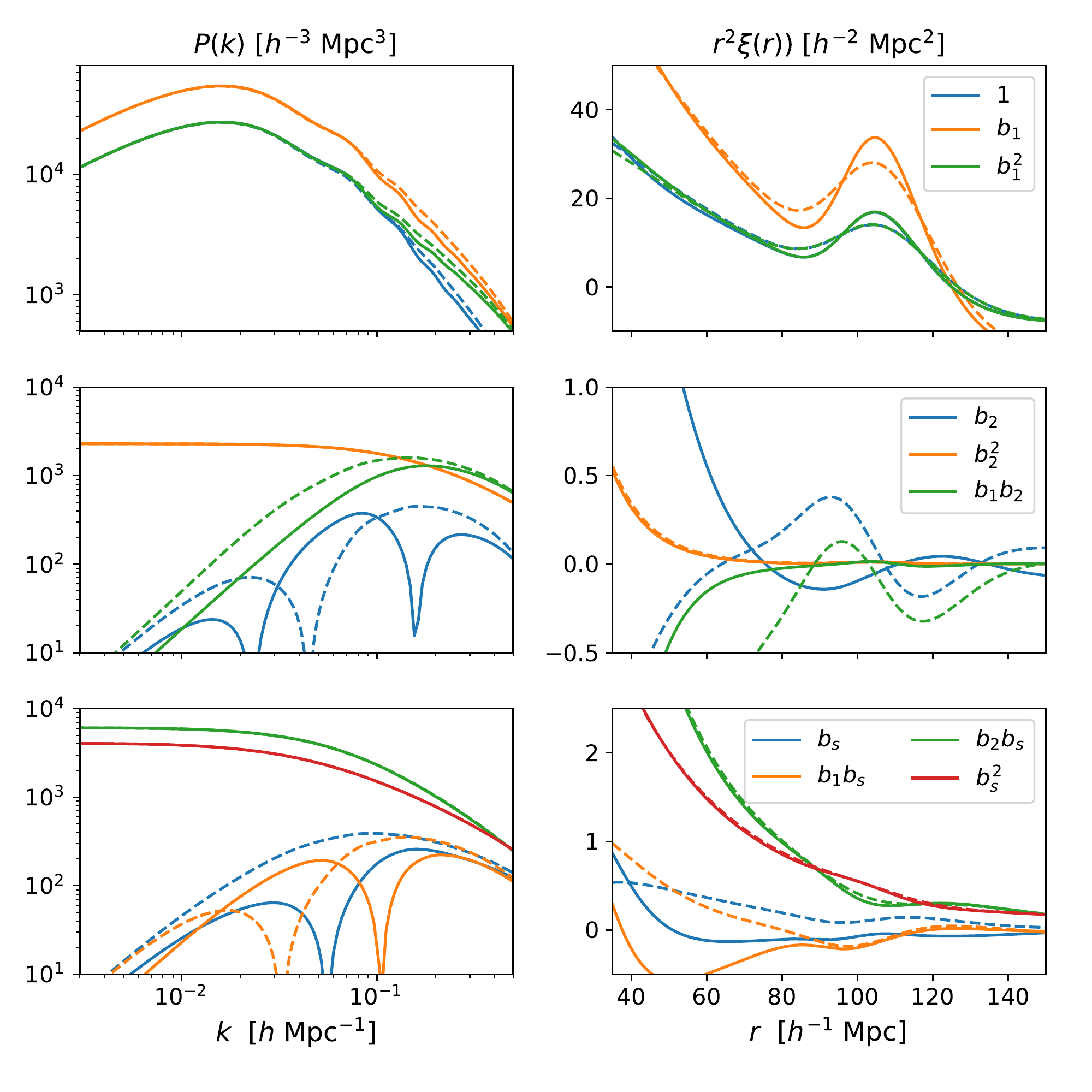}
    \caption{Contributions to the pre- and post-reconstruction (dashed and solid) power spectra and correlations functions (left and right columns) in real space from linear through quadratic bias terms at $z = 0$. Note that the matter (blue) and $b_1^2$ (green) curves in the top right panel are essentially degenerate, especially at the large scales shown. }
    \label{fig:precon_biasterms_z0}
\end{figure}

The lowest-order bias terms in the reconstructed real-space power spectrum at $z = 0$ are shown Figure~\ref{fig:precon_z0}. The pure-matter piece (i.e. the ``1'' in Equation~\ref{eqn:cross_spectra}) is the only term that includes contributions from all three combinations of $d$ and $s$, while the $b_1^2$ piece consists of only the $dd$ contribution. While each piece individually differs from the linear power spectrum, compared to the pre-reconstruction power spectrum, the Zeldovich approximation predicts that the post-reconstruction power spectrum largely recovers the oscillatory features in the linear spectrum, as seen in the lower panels of Figure~\ref{fig:precon_z0}. We note that the structure of the breakdown into $P^{dd}$, $P^{ds}$ and $P^{ss}$ shown in Figure~\ref{fig:precon_z0} proceeds similarly in the higher-order bias contributions: bias terms like $b_1^2$, that are products of two bias parameters (e.g.\ $b_1 b_2, b_2 b_s$, ...), do not involve any displacements ($\bPsi$) and can thus only enter in the autospectrum of the biased ``$d$'' tracer $P^{dd}$, while those like $b_1$ that involve only one bias parameter (e.g.\ $b_2$, $b_s$) involve two-point functions with one displacement contracted and thus contribute to the cross spectrum $P^{ds}$ as only one of the constituent tracers needs to be biased. The autospectrum $P^{ss}$ does not contain any bias terms.

Figure~\ref{fig:precon_biasterms_z0} shows all the contributions to the reconstructed galaxy power spectrum and correlation function, up to the quadratic bias and shear terms. As seen in the top panel, the reduced damping in the lowest-order bias term ``wiggles,'' barely visible in log-log plots of the reconstructed power spectrum, translate to significantly sharper and less shifted BAO features (right column). In the quadratic bias contributions (middle panels), reconstruction can be seen to dampen the amplitude of the BAO feature in the $b_2$ and $b_1 b_2$ contributions, which ``wiggle'' in Fourier space, while leaving the spectrally smooth $b_2^2$ contribution essentially intact.  Since the BAO feature in the quadratic bias contributions will tend to smear and shift the observed BAO peak from its linear theory position, reconstruction serves to remove these confounding nonlinearities as expected. The shear terms have less pronounced (i.e. smoother) features at the BAO scale---which we will show in Section~\ref{sec:earlier} as being essentially in-phase with the linear theory oscillations--- that are less affected by reconstruction.

Finally, as noted in the discussion below Equation~\ref{eqn:cross_spectra}, the exponentiated $A_{ij}$ in Zeldovich power spectra are assumed to be long wavelength, IR modes which can be resummed while contributions from the rest of the shorter modes are perturbatively expanded. These expanded modes thus carry also a UV (small-scale) sensitivity that should be renormalized by adding the appropriate counterterms, quadratic in wavenumber and proportional to the Zeldovich power spectrum: $\alpha_{ab} k^2 P_{\rm Zel}^{ab}$.  In principle, we expect such counterterms in all three pieces of our reconstructed power spectrum, however given that the $P^{ss}$ consists of mostly IR modes we expect its counterterm contribution to be suppressed relative to similar terms in $P^{dd}$ and $P^{ds}$, though it could still be non-vanishing due to contributions we neglected in approximating Equation~\ref{eqn:recon_shift}. While the counterterms $\alpha_{dd}$ and $\alpha_{ds}$ are highly nondegenerate due to the different supports of $P^{dd}$ and $P^{ds}$ (see Figure~\ref{fig:precon_z0}) in $k$-space, in this work we will also explore modelling the reconstructed power spectrum using only one counterterm, $\propto k^2 P_{\rm Zel}$, for both $P^{dd}$ and $P^{ds}$ contributions, since such a contribution would also be degenerate with any potential derivative biases (see e.g.\ ref.~\cite{VlaCasWhi16}). We will return to the difference between these options in Section~\ref{sec:nbody}.

\subsection{Redshift space}
\label{sec:redshift_space}

In this section we develop analytic expressions for the redshift-space reconstructed power spectrum in both \textbf{Rec-Sym} and \textbf{Rec-Iso}.
Methods recently developed in ref.~\cite{VlahWhi18} allow us to extend the LPT redshift-space power spectrum calculation to include bias and the specifics of reconstruction, which we summarise here and present in detail in Appendix~\ref{sec:rsd_integrals}. As we will show shortly, \textbf{Rec-Sym} and \textbf{Rec-Iso} are not equivalent even to linear order. Specifically, we have
\begin{align}
    &P_{\rm sym}(\bk) = (b+f\nu^2)^2 P_{\rm L}(k) + \mathcal{O}(P_{\rm L}^2) \\
    &P_{\rm iso}(\bk) = \Big[(b+f\nu^2) (1 - \mathcal{S}) + b\ \mathcal{S} \Big]^2 P_{\rm L}(k) + \mathcal{O}(P_{\rm L}^2),
\end{align}
i.e.\ while \textbf{Rec-Sym} restores supercluster infall at linear order, \textbf{Rec-Iso} removes redshift-space distortions at large scales while keeping them at small scales\footnote{We have amended Equation~4.11 to correct for a typo pointed out in ref.~\cite{Philcox20}, wherein the $b\mathcal{S}$ in Equation~4.11 was missing a factor of $b$. No other results or conclusions are affected.}. As we will see, this produces a smooth modulation in the broadband power nondegenerate with the BAO wiggles.

Since both the smoothed and displaced fields are uniformly multiplied by $R_{ij}$ in \textbf{Rec-Sym}, it is straightforward to calculate the reconstructed power spectrum using Equation~\ref{eqn:cross_spectra} with
\begin{equation}
    \bPsi^d(\bk) = \left[1 - \mathcal{S}(k)\right]\mathbf{R}\bPsi(\bk) \quad , \quad \bPsi^s(\bk) = - \mathcal{S}(k)\mathbf{R} \bPsi(\bk) \quad .
\end{equation}
In particular the angular structure of the $\bq$ integral follows as in the calculation of the galaxy power spectrum without further modifications, and the set of bias terms in the $dd$, $ds$ and $ss$ spectra are identical to the real space case. The reconstructed power spectrum can then be calculated as one would the unreconstructed redshift space power spectrum. We develop the formalism to do the latter in Appendix~\ref{sec:MII} and comment on the changes required to go to the reconstructed case therein.

The cross spectrum in \textbf{Rec-Iso} is slightly different since only the displaced field is multiplied by the redshift space transformation, $R_{ij}$.   The displaced and shift fields in this case are thus instead
\begin{equation}
    \bPsi^d(\bk) = \left[1 - \mathcal{S}(k)\right]\bPsi^R(\bk) = \left[ 1 - \mathcal{S}(k)\right]  \mathbf{R} \bPsi(\bk) \quad , \quad \bPsi^s(\bk) = - \mathcal{S}(k) \bPsi(\bk).
\end{equation}
Since the displaced and shift moves thus lie in redshift and real space, respectively, their auto spectra can also respectively be calculated as in \textbf{Rec-Sym} and real space reconstruction; however, the cross spectrum is only ``half transformed'' into redshift space and thus requires special attention.
The exponentiated two-points displacements are given by
\begin{align}
    A_{ij}^{ds, \rm Iso} &= \langle \bPsi^d_i \bPsi^d_j \rangle +  \langle \bPsi^s_i \bPsi^s_j \rangle -  2\langle \bPsi^d_i(\bq_2) \bPsi^s_j(\bq_1) \rangle \nonumber \\
    &= R_{in} R_{jm} \langle \bPsi^d_n \bPsi^d_m \rangle_{\rm Real Space} +  \langle \bPsi^s_i \bPsi^s_j \rangle_{\rm Real Space} -2  R_{in} \langle \bPsi^d_n(\bq_2) \bPsi^s_j(\bq_1) \rangle_{\rm Real Space},
    \label{eqn:iso_aij}
\end{align}
such that the zero-lag piece due to the displaced-displaced correlation is fully transformed into redshift space, the zero-lag piece due to the shifted-shifted correlation is untransformed, and the coordinate dependent displaced-shifted correlation is``half transformed.''
In particular, defining as usual $\bq = q\ \hq$ and $\hat{k} \cdot \hat{q} = \mu$, the last piece is
\begin{align}
    k_i k_j \langle \bPsi^d_i(\bq_2) \bPsi^s_j(\bq_1) \rangle = k_i k_j (\delta_{ik} &+ f \hat{n}_i \hat{n}_k) (\tilde{X}^{ds} \delta_{kj} + \tilde{Y}^{ds} \hq_k \hq_j) \nonumber \\
    &= k^2 (1 + f\nu^2) \tilde{X}^{ds} + k^2 (\mu^2 + f\mu \nu (\hq \cdot \hat{n})) \tilde{Y}^{ds},
\end{align}
where we have defined the tilded quantities without the usual zero lag piece\footnote{For notational simplificity, the functions $X$ and $Y$ are always defined in real space.}
\begin{equation}
    \langle \bPsi^{d}(\bq_2) \bPsi^{s}(\bq_1) \rangle_{\rm Real\ Space} = \tilde{X}^{ds}(q)\ \delta_{ij} + \tilde{Y}^{ds}(q)\ \hq_i \hq_j. \nonumber
    \label{eqn:nozlag}
\end{equation}
Note that $2\tilde{Y} = -Y$ since $Y$ does not posess a zero-lag piece. The azimuthal-angle dependence in $\hq \cdot \hat{n}$ will require us to do the integral (Appendix~\ref{sec:MI})
\begin{equation}
    \int \frac{d\phi}{2\pi} e^{A \mu \sqrt{1-\mu^2} \cos\phi} = \sum_{\ell=0}^{\infty} H^{(0)}_\ell(A)\ (A\mu^2)^\ell, \nonumber
\end{equation}
where we have defined
\begin{equation}
    H^{(0)}_\ell(A) = \sum_{m=0}^\ell 
    \frac{(-1)^{\ell-m}  A^{2m-\ell} \Gamma(m+\frac{1}{2})}{\sqrt{\pi}\Gamma(2m+1)\Gamma(2m-\ell+1)\Gamma(\ell-m+1)}. \nonumber
\end{equation}
Note that the $\Gamma$ functions in the denominator will kill any terms in the sum for which $2m-\ell$ is negative, such that the sum really only contains $\ell/2$ terms and is always convergent in $A$. The full cross spectrum is then given by 
\begin{align}
    P^{(ds)}(\bk) =  &e^{-\frac{1}{2}k^2 (\alpha_{0} \Sigma^{(dd)^2} + \Sigma^{(ss)^2})}  \dtq e^{ikq\mu + k^2 (1 + f\nu^2)(\tilde{X}^{(ds)} + \mu^2 \tilde{Y}^{(ds)})} \nonumber \\
    &\sum_{\ell=0}^{\infty} H^{(0)}_{\ell}(A) A^{\ell} \mu^{2\ell} \Bigg(1 + i b_1 k\mu U^{(d)}(q) - \frac{1}{2} b_2 k^2 \mu^2 U^{(d)}(q)^2 + ... \Bigg)
    \label{eqn:methodii_cross}
\end{align}
where $A = k^2 f\nu\sqrt{1-\nu^2}\tilde{Y}^{(ds)}$ and we have defined $\Sigma^2 = \tilde{X}(0)$ and $\alpha_0 = 1 + f(f+2)\nu^2$. The remaining integrals can then be performed using the usual tricks for powers of $\mu$ using the series described in Appendix~\ref{sec:mu_ints}, and are explicitly given at the end of Appendix~\ref{sec:MI}.

Figures~\ref{fig:precon_biasterms_z0_sym} and \ref{fig:precon_biasterms_z0_iso} show the various bias contributions to the reconstructed redshift space power spectrum monopoles and quadrupoles within \textbf{Rec-Sym} and \textbf{Rec-Iso}, respectively. A significant difference between the two methods can be seen by comparing the matter (i.e.\ ``1'') pieces in the top panels of the two figures. While all three linear bias contributions to the reconstructed power spectrum monopole ($\propto 1$, $b_1$, $b_1^2$) approach the Zeldovich monopole in the large scale limit in \textbf{Rec-Sym}, the matter contribution to the \textbf{Rec-Iso} monopole instead approaches the $b_1^2$ contribution, which does not receive redshift space distortions in the linear theory limit. This is because the power spectrum at the largest scales is dominated by the autospectrum of the un-redshifted shift field, $P^{ss}$. While the matter and $b_1$ contributions to the reconstructed quadrupole approach linear theory in \textbf{Rec-Sym}, they vanish on large scales in \textbf{Rec-Iso}. On the other hand, the majority of the higher bias contributions (excluding $b_2$ and $b_s$) are sourced only by $P^{dd}$ and are thus identical between the two methods, as can be seen by comparing the lower two rows of Figures~\ref{fig:precon_biasterms_z0_sym} and \ref{fig:precon_biasterms_z0_iso}.  This corresponds to our intuition that redshift-space distortions are less prominent for highly biased tracers, and that the differences between Rec-Iso and Rec-Sym disappear if we remove RSD. In addition, the contributions enumerated above are supplemented by counterterms $(\alpha^{\ell}_{dd},\alpha^{\ell}_{ds},\alpha^{\ell}_{ss})$, where we need a separate counterterm for each pair and multipole as discussed below Equation~\ref{eqn:zerolag_rsd}, though as in the real space case we also explore the possibility of only fitting one counterterm each for the net reconstructed monopole and quadrupole.

\begin{figure}
    \centering
    \includegraphics[width=\textwidth]{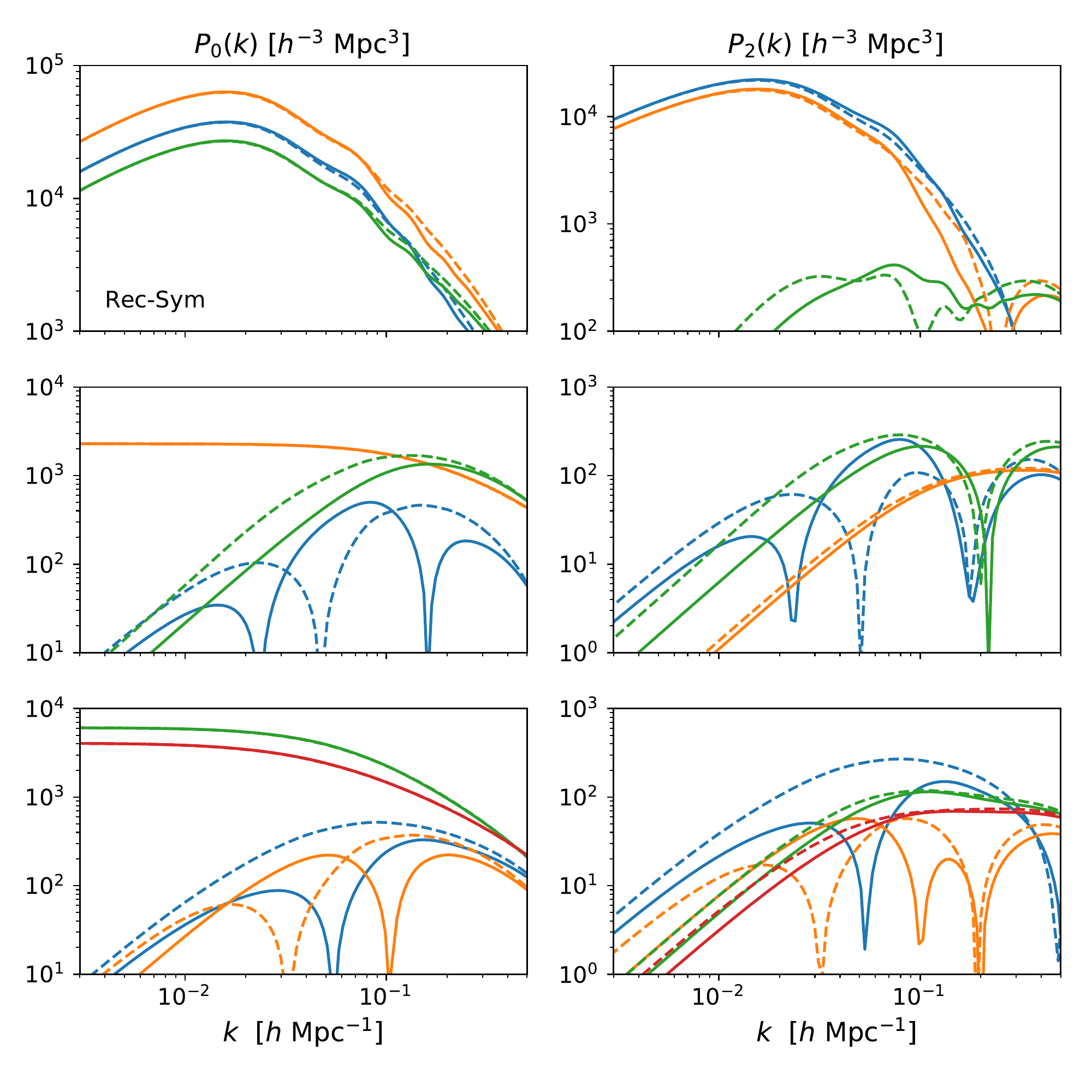}
    \caption{Bias contributions to the pre- and post-reconstruction (dashed and solid) $z = 0$ redshift space power spectra monopole and quadrupoles in the \textbf{Rec-Sym} scheme. The color scheme and line styles follow those in Figure~\ref{fig:precon_biasterms_z0}. The lowest-order contributions to the reconstructed monopole and quadrupole due to the linear bias $b_1$ tend to the Kaiser approximation at large scales.  Note the different $y$-axis ranges on different panels.}
    \label{fig:precon_biasterms_z0_sym}
\end{figure}

\begin{figure}
    \centering
    \includegraphics[width=\textwidth]{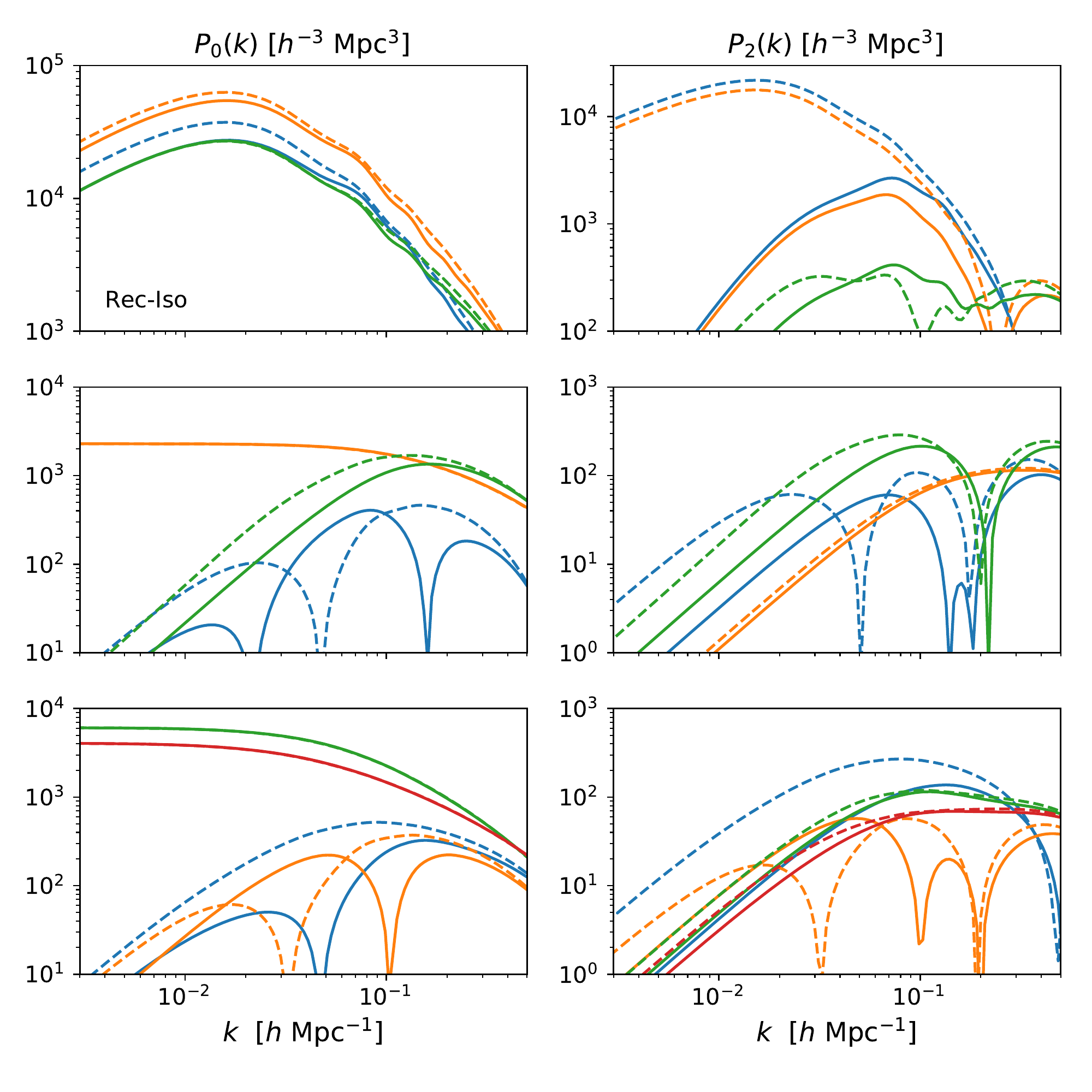}
    \caption{Same as Figure~\ref{fig:precon_biasterms_z0_sym}, but for \textbf{Rec-Iso} at $z = 0$. Unlike in \textbf{Rec-Sym}, the linear bias contributions to the monopole and quadrupole do not tend to the Kaiser limit on large scales but to the real space linear power spectrum, as evidenced by reduced power in the monopole compared to the pre-reconstruction Zeldovich power spectrum, and contributions to the quadrupole vanishing on large scales. However, many of the higher bias contributions are identical to those in \textbf{Rec-Sym} (Fig.~\ref{fig:precon_biasterms_z0_sym}).}
    \label{fig:precon_biasterms_z0_iso}
\end{figure}

\section{Reconstructed correlation function}
\label{sec:config}

The configuration space two-point function (the correlation function) can be obtained from our Fourier-space results by Hankel transform.  It is also possible to rewrite the $q$-dependent integrals to compute $\xi(r,\nu_r)$ directly, where $\nu_r = \hat{n}\cdot\hat{r}$.  
Here we reprise the calculation of ref.~\cite{Whi15}, extending it to include the additional bias terms and commenting explicitly on several numerical issues which arise. We have checked that our Fourier and configuration space results agree numerically to significantly sub-percent levels in both real and redshift space for both \textbf{Rec-Sym} and \textbf{Rec-Iso}.

The general formula for the cross spectrum of two tracers $a$ and $b$ given in Equation~\ref{eqn:cross_spectra} can be Fourier transformed to give \cite{Whi14,VlaCasWhi16,VlaWhiAvi15,WanReiWhi14}
\begin{eqnarray}
    1+\xi^{ab}(r) &=& \int 
    \dfrac{d^3q}{(2\pi)^{3/2}|A^{ab}|^{1/2}}
    e^{-(1/2)(q_i-r_i)(A^{-1}_{ab})_{ij}(q_j-r_j)}
    \nonumber \\
    &\times& \bigg\{ 1 - 
    (b_1^b U^a_i + b_1^a U^b_i) g_i
    + b_1^a b_1^b \xi_L + \dfrac{1}{2}b_2^a b_2^b \xi_L^2
    \nonumber \\
    &&- \dfrac{1}{2}\ [ b_2^b U^a_i U^a_j + b_2^a U^b_i U^b_j  + 2 b_1^a b_1^b U_i^a U_j^b ]\ G_{ij}
    - [b_1^a b_2^b U_i^a + b_1^b b_2^a U_i^b]\ \xi_L g_i
    \nonumber \\
    &&- \frac{1}{2}\ [b_s^a \Upsilon^b_{ij} + b_s^b \Upsilon^a_{ij}] \ G_{ij}
    -[b_1^a b_s^b V_i^{ab} + b_1^b b_s^a V_i^{ba} ]\ g_i \nonumber \\
    &&+ \frac{1}{2}(b_2^a b_s^b + b_2^b b_s^a) \chi^{12} + b_s^a b_s^b\zeta + \alpha_{ab} \text{tr}G +  ...
    \bigg\} \ ,
    \label{eqn:gen_correlation}
\end{eqnarray}
where we have defined
\begin{equation}
    g_i = (A_{ab}^{-1})_{ij}(q_j - r_j), \; G_{ij} = (A_{ab}^{-1})_{ij} - g_i g_j
\end{equation}
and placed the superscript $ab$ in $A^{ab}$ into the subscript for notational convenience.
In the configuration space calculation above, the Lagrangian two-point functions (e.g.\ $A_{ij}$, $U_i$, $\Upsilon_{ij}$) can be computed using the formulae provided in Appendix~\ref{sec:correlators}.  The above formula can be translated into redshift space by multiplying the Lagrangian two-point functions with vector indices by the appropriate factors of $R_{ij} = \delta_{ij} + f \hat{n}_i \hat{n}_j$. Taking the line-of-sight to be in the $z$ direction without any loss of generality, this is equivalent to multiplying by the matrix $\text{diag}(1,1,1+f)$. When calculating the un-reconstructed redshift space correlation function, this multiplication is equivalent to multiplying each $z$ component index of vector and tensor quantities (e.g.\ $U^a_z$ or $A^{ab}_{yz}$) by $1+f$, and dividing the corresponding components in the matrix inverse, $A_{ab}^{-1}$, by the same factor. The redshift-space counterterm $\alpha_2 k^2 \nu^2$ can be included in the correlation function by adding $\alpha_2 \hat{n}_i \hat{n}_j G_{ij},$ which similarly is equivalent to $\alpha_2 G_{zz}$ when picking $z$ as the line-of-sight direction. 

The reconstructed correlation function in real and redshift space can be calculated using Equation~\ref{eqn:gen_correlation} by defining ``displaced'' and ``shifted'' tracers as in the case of the power spectrum (Sections~\ref{sec:pk_real} and \ref{sec:redshift_space}) and calculating the combined quantity $\xi_{\rm recon} = \xi_{dd} + \xi_{ss} - 2 \xi_{ds}$.  For reconstruction using \textbf{Rec-Sym}, the same shortcuts of multiplying by factors of $1+f$ in lieu of matrix multiplication and inversion apply, since all vector and tensor quantities undergo the same transformation by $R_{ij}$. The calculation for \textbf{Rec-Iso} is more complicated. As was the case in Fourier space, the displaced-displaced and shifted-shifted auto-correlation functions are equal to their counterparts in \textbf{Rec-Sym}
and real-space reconstruction, while the displaced-shifted cross-correlation function contains a mix of real and redshift space factors. In particular, from Equation~\ref{eqn:iso_aij} we see that the two zero-lag pieces and one $q$-dependent piece of $A^{ds}_{ij}$ in \textbf{Rec-Iso} are independently transformed by different numbers of $R_{ij}$'s. For this reason, when calculating the correlation function in \textbf{Rec-Iso}, the matrix inverse of $A^{ds}$ in redshift space cannot be simply obtained by dividing the real space inverse by factors of $1+f$; rather, the uninverted matrix must be redshifted piece by piece as in Equation~\ref{eqn:iso_aij} and then inverted numerically (we use Cholesky decomposition).

\section{Other statistics}
\label{sec:omega}

While the correlation function and power spectrum are the most frequently considered 2-point functions, there are other variants that have some advantages.  Since these can all be written in terms of the correlation function or power spectrum, our model provides a consistent prediction for them as well.  Of particular interest for BAO is the $\omega_\ell$ statistic of ref.~\cite{Xu10}, which combines the scale-localization of the Fourier-space methods with the compactness and easy treatment of masks of the configuration-space methods.

In principle $\omega_\ell$ can be calculated from either the configuration-space or Fourier-space expressions given above, but we have found it more convenient to start from the Fourier expressions.  Since these are computed using FFTlog they naturally cover a very wide range of $k$, making the transforms to $\omega_\ell$ easy to implement.  For example
\begin{equation}
    \omega_0(r_s) = \int\frac{k^2\,dk}{2\pi^2} P_0(k) \widetilde{W}_0(k\,r_s) 
\end{equation}
with $\widetilde{W}_0$ given in ref.~\cite{Xu10} (see their Fig.~1 and Appendix A).  At large scales $\widetilde{W}_0\propto k^2$ while at small scales $\widetilde{W}_0\propto k^{-4}$.  Our formalism naturally provides predictions for $\omega_\ell$ using the same set of bias and nuisance parameters as for $\xi_\ell$ and $P_\ell$.

\section{Comparison to N-body}
\label{sec:nbody}

\begin{table}
\begin{center}
\begin{tabular}{cccc}
${\rm lg}M$ & Redshift & $\bar{n}$ & $b$ \\ \hline
$12.0-12.5$ &   0.0    & $3.45$    & 0.87  \\
$12.5-13.0$ &   0.0    & $1.18$    & 1.05 \\
$13.0-13.5$ &   0.0    & $0.38$    & 1.30
\end{tabular}
\caption{Number densities and bias values for the halo samples we use.
Halo masses are $\log_{10}$ of the mass in $h^{-1}M_\odot$, number densities
are times $10^{-3}\,h^3{\rm Mpc}^{-3}$.}
\label{tab:nbody}
\end{center}
\end{table}

To look at the domain of validity of our analytic results we compare to the {\tt DarkSky} N-body simulation suite\footnote{http://darksky.slac.stanford.edu}, specifically simulation {\tt ds14\_a} \cite{skillman14}.  This simulation used the {\tt 2HOT} code \cite{War13} to evolve $10240^3$ particles in an $(8\,h^{-1}$Gpc$)^3$ volume to model the growth of structure in a $\Lambda$CDM cosmology with $\Omega_M=1-\Omega_\Lambda=0.295$, $h=0.688$,
$n_s=0.968$ and $\sigma_8=0.835$.  Initial conditions were generated from a glass using $2^{\rm nd}$ order Lagrangian perturbation theory at $z=93$.  Halos were found using the {\tt Rockstar} code \cite{Beh13}.  We extracted the positions, velocities and masses of halos more massive than $M_{200b}=10^{12}\,h^{-1}M_\odot$ from the publicly available data at $z=0$ (data at higher $z$, which would have been a more relevant comparison, were not available). We computed the halo correlation functions and power spectra, in real and redshift space.  For the redshift-space quantities we assumed the plane-parallel approximation with the line-of-sight being the $z$-axis. We also obtained the linear theory power spectrum used to generate the initial conditions, which we take as the input to our model.

We implemented the algorithm described in \S\ref{sec:recon} using the periodicity of the box and FFTs to perform the smoothing and computation of the shifts.  As for the power spectrum and correlation function, the plane-parallel approximation with line-of-sight the $z$-axis was assumed for the redshift-space quantities.  The code takes as input an assumed large-scale bias, $b$, and growth parameter, $f$, in addition to a Gaussian smoothing length, $R$.  We used the $b$ obtained from the ratio of the linear theory and real-space halo power spectra at low $k$ (see Table \ref{tab:nbody}), and $f\simeq 0.508$ appropriate to the simulation cosmology at $z=0$, and note in passing that the goodness-of-fit of our results did not seem to be greatly improved by substituting the linear bias thus obtained with the value of $1 + b_1$ obtained by fitting the pre-reconstruction data with our model up to quasi-nonlinear scales.

We computed the reconstructed field in both real and redshift space.  In each case the shifted and displaced positions were computed using a $2048^3$ FFT, which resolves the (Gaussian) smoothing length by $2.5-5$ grid cells for $R\simeq 10-20\,h^{-1}$Mpc.  We used as many ``random'' positions as halos in each case, for simplicity, and computed the power spectra and correlation functions for $dd$, $ds$ and $ss$ assuming periodic boundary conditions.  The reconstructed power spectrum or correlation function can then be computed as $dd-2ds+ss$, and we can look at each of the contributions separately.  Note that our choice of equal numbers of randoms and data points means the shot noise on the reconstructed power spectrum is twice that of the pre-reconstructed field.

\begin{figure}
    \centering
    \includegraphics[width=0.95\textwidth]{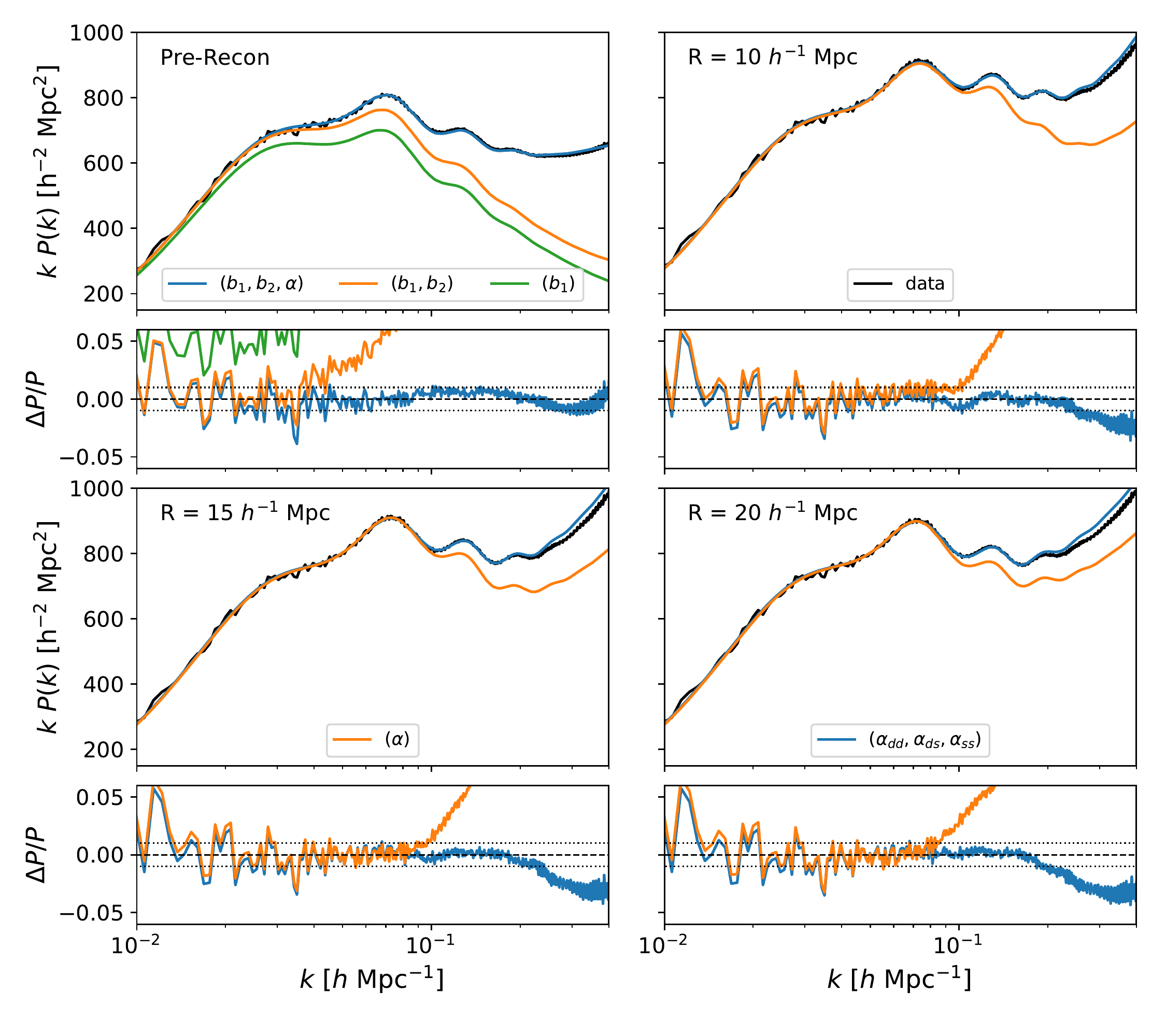}
    \caption{Fits to the pre- and post-reconstruction real-space halo power spectra in {\tt DarkSky} for halos of mass between $12.5 < \log_{10}(M/ h^{-1} M_\odot) < 13.0$ at three smoothing scales ($R = 10$, $15$, $20\kMpc$), assuming Zeldovich power spectra with biases $(b_1, b_2)$ and one counterterm per spectrum (three total for the reconstructed case). The upper plot of each vertical pair of panels shows the product of the wavevector magnitude and power spectrum $k\ P(k)$ while the lower plot shows the fit residuals as a fraction of measure power $\Delta P/P = (P_{\rm fit} - P_{\rm nbody})/P_{\rm nbody}$. In the top-left pair of panels we show the incremental contributions from $b_2$ and the counterterm $\alpha$ (which contributes close to $10\%$ of the power at $k = 0.1 \kMpc$) to the fit, which agrees with the simulation at the percent level (dotted line in the lower plots) at all scales shown. In the remaining panels we use the same bias parameters to fit the reconstructed power spectrum, allowing only counterterms to vary. Our model with three counterterms can fit the data at the percent level out to $k = 0.2 \kMpc$, though a bump-like feature at $k = 0.1 \kMpc$ becomes more prominent at smaller smoothing scales, where nonlinear corrections beyond the Zeldovich approximation presumably become more important (see text). Also shown in orange are fits using one counterterm -- or equivanlently one derivative bias -- which fit less well past $k = 0.1 \kMpc$. We fined that setting the counterterm $\alpha_{ss}$ to zero does not materially affect our fits. Note that there is excess power in the data at the largest scales shown, as discussed in the text.
    }
    \label{fig:real_space_pks}
\end{figure}

We compare the N-body results to our model with $b_1$ and $b_2$ and include the minimal set of counterterms as described in the preceding sections (one and three pre- and post-reconstruction, respectively, in real space) as well as a constant shot noise component fit to the data. For brevity our discussion will focus on halos with masses between $12.5 < \log_{10}(M/ h^{-1} M_\odot) < 13.0.$, though we obtained qualitatively similar results in the lower and higher mass bin as well, and show fits of the reconstructed redshift space power spectrum in the latter at the end of this section. We have checked that including nonzero shear bias $b_s$ does not visibly improve the goodness-of-fit. The top-left pair of panels of Figure~\ref{fig:real_space_pks} compares the unreconstructed real-space power spectrum in our model with $(b_1, b_2) = (0.02, -0.8)$ with that in {\tt DarkSky}. The quadratic bias, $b_2$, accounts for a non-negligible fraction of the total power at essentially all scales and significantly reduces the constant shot noise term in the fit. We find that with a counterterm $\alpha \approx 11\ h^{-2} \text{Mpc}^2$ our model agrees with the data at the percent level out to $k\simeq 0.4\kMpc$.  The counterterm accounts for roughly a $10\%$ correction at $k = 0.1\kMpc$, and it is worth noting that even in its absence our model accurately captures the BAO features in the power spectrum, as evidenced by the lack of oscillatory features in the fit residuals.

The remaining panels of Figure~\ref{fig:real_space_pks} show the fit for the reconstructed power spectra at three smoothing scales $R = 10$, $15$, $20\,h^{-1}$Mpc. 
We have tested whether the data could be reproduced using only one counterterm, $\alpha$ (shown in orange), or equivalently from one derivative bias $b_{\nabla^2}$, and find that such a choice dramatically reduces the range-of-validity of the model compared to three counter terms.  While we adopted a rather conservative approach in fitting these data, prioritizing the accuracy of our predictions at low $k$ rather than producing reasonable-looking fits to smaller scales, our model with three counterterms $(\alpha_{dd},\alpha_{ds},\alpha_{ss})$ nonetheless reproduces both the broadband power and oscillatory features of the reconstructed power spectrum out to $k = 0.2 \kMpc$ at the percent level for $R = 15$ and $20 h^{-1}$ Mpc. That each of the three constituent spectra in $P^{\rm recon} = P^{dd} + P^{ss} - 2 P^{ds}$ has distinct short-wavelength behavior and $k$-space supports underlies the success of our model with three counterterms---each of which has highly nondegenerate scale dependence---versus the one-counterterm alternative. We have found that setting $\alpha_{ss} = 0$ does not qualitatively alter the degree to which our model fits the data; we have made this choice in all of our fits below, but note that as $\alpha_{ss} k^2$ vanishes quadratically towards low $k$, the data are also naturally rather insensitive to it. Indeed, since nonlinear corrections are typically of order $k^2 \Sigma^2$, and the smoothing scale is chosen such that $(\Sigma/R)^2 \ll 1$, the insensitivity of $P^{ss}$ to these corrections follows almost by construction. However, a bump-like feature around $k = 0.1\kMpc$ is persistent across all the fits, peaking at less than half a percent when $R = 20\,h^{-1}$Mpc and growing to a full percent at $R = 10\,h^{-1}$Mpc. The appearance of such a feature, growing towards smaller smoothing scales, is consistent with our neglect of nonlinear corrections to the smoothed displacements, which should increase towards smaller smoothing scales roughly as $\bPsi/R$; we discuss one such nonlinearity in Appendix~\ref{app:EulerianPosition}. For sufficiently small smoothing scales, even the assumption that the smoothing of the BAO feature can be essentially captured with resummed \textit{linear} displacements $\bPsi^{d,s}$ will break down, and indeed our fit residuals begin to show noticeable oscillatory behavior at the smallest smoothing scale shown ($R = 10\,h^{-1}$Mpc). At $R = 15\,h^{-1}$Mpc and in the sample variance limit with Gaussian errors, the feature at $k=0.1\kMpc$ should be detectable with $\chi^2 = V_{\rm obs} / (2 h^{-3}\ \text{Gpc}^3 )$, where $V_{\rm obs}$ is the total observed volume. If we were to instead smooth using the larger $R = 20\,h^{-1}$Mpc, the $\chi^2$ is roughly halved.  For such a smoothing this feature represents a $\chi^2$-penalty of $0.2$ for a sample variance limited survey of $14\,000\,\mathrm{deg}^2$ covering $0\le z\le 0.3$, and would be slightly smaller for finite number density.

\begin{figure}
    \centering
    \includegraphics[width=\textwidth]{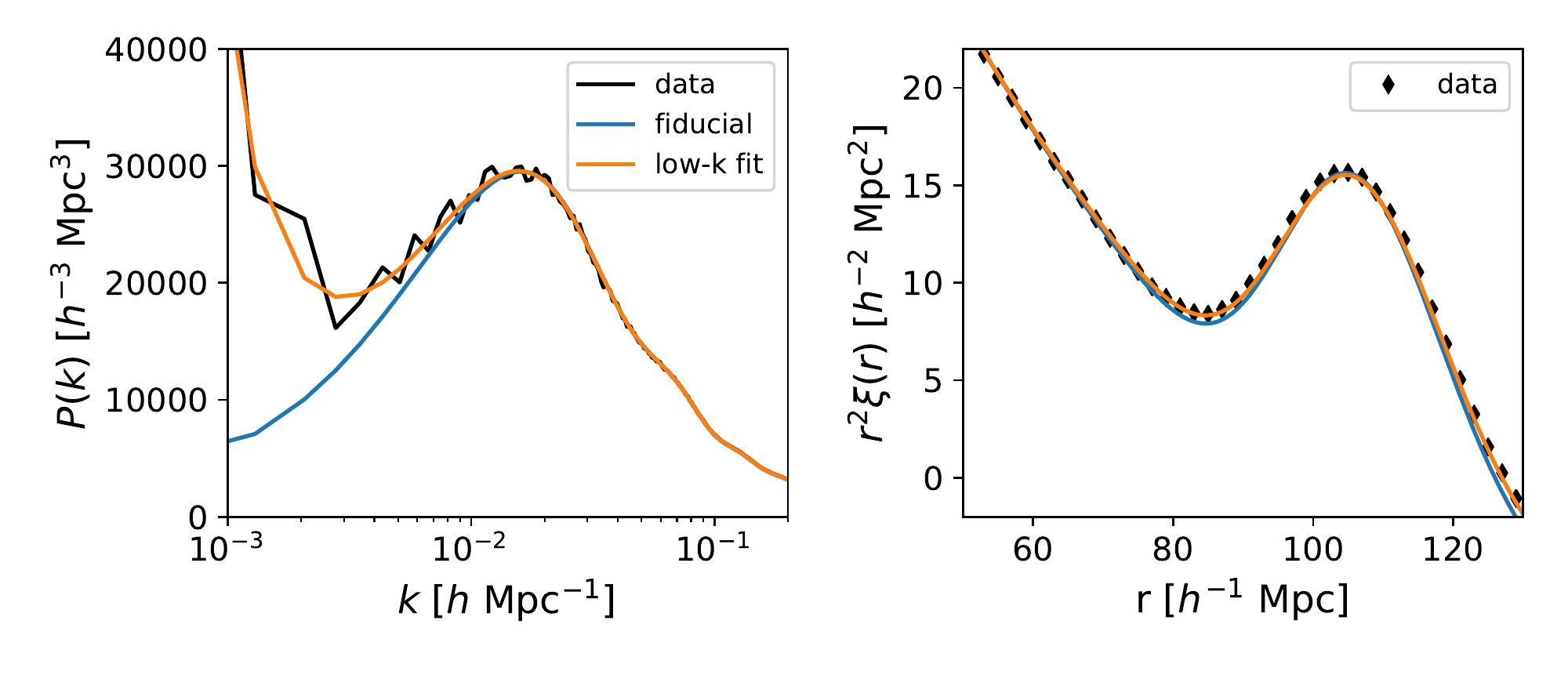}
    \caption{Halos in {\tt DarkSky} exhibit significant excess power compared to theory at large scales in Fourier space which should be well-described by linear theory. (Left) Fits to the real-space power spectrum with and without our ad hoc correction $P_{\rm lw} = A\ (k/k_0)^n,$ shown in blue and orange respectively. At the largest scales shown, the excess power is significantly larger than the scatter. The fits prefer slighly different, though qualitatively similar, bias values.  (Right) The same fits in configuration space. The uncorrected data systematically trends below the data at separataions above the BAO peak and in the BAO ``dip,'' while the fit with $P_{\rm lw}$ added goes through all the data points.}
    \label{fig:lowk_fit}
\end{figure}

\begin{figure}
    \centering
    \includegraphics[width=\textwidth]{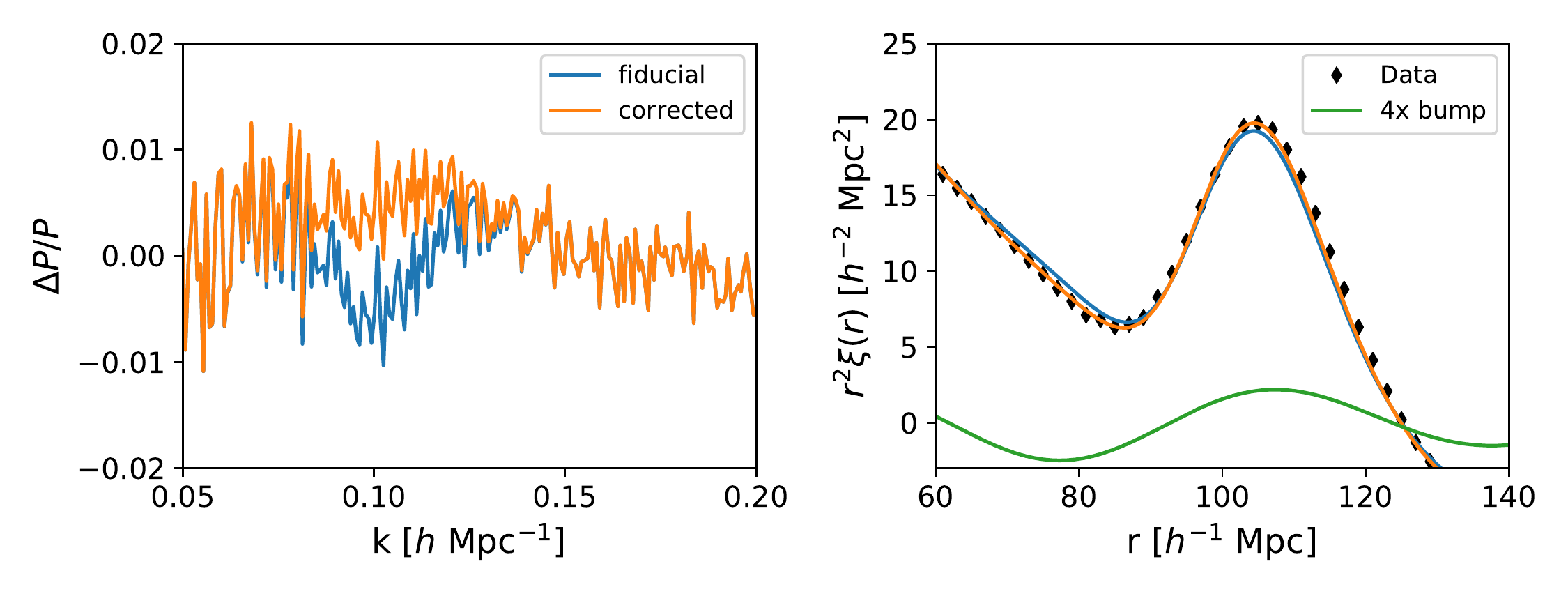}
    \caption{A sub-percent level feature in the power spectrum near $k=0.1\kMpc$ can lead to visible distortions in the BAO feature in $\xi(r)$. (Left) Residuals for the fit as a fraction of total measured power in the simulations, as defined in the caption of Figure~\ref{fig:real_space_pks}. The orange curve shows the residuals when our theory is corrected using a Gaussian profile localized at $k = 0.1 \kMpc$ compared to the fiducial fit (blue), whose residuals exhibit a dip centered at $k = 0.1 \kMpc.$ (Right) The fiducial and corrected correlation functions. The bump in the left panel, whose Fourier transform is shown magnified in the green curve, induces distortions in the BAO feature across a range of separations $r \sim 60 - 120\,h^{-1}$Mpc.
    }
    \label{fig:ft_bump}
\end{figure}

The pre- and post-reconstruction real-space correlation functions can be directly compared by computing the Fourier transforms of the above fits. However, in comparing our theory with {\tt DarkSky} we found that the $z=0$, pre-reconstruction halo power spectra all have significant excess power at low $k$ compared to the predictions of linear theory with scale-independent bias.  The origin of this excess is unclear, and is not addressed in ref.~\cite{skillman14}.  It appears to arise from a significant number of low $k$ modes, and so is unlikely to be simply a statistical fluctuation in the initial conditions.  It shows up in all of our halo samples, and is highly correlated among mass bins.  This excess power is small for modes to the right of the power spectrum peak and probably has only a small impact on the dynamics on BAO scales.  In Fourier space we simply confine our fitting and modeling to $k>0.01\,h\,\mathrm{Mpc}^{-1}$.  In configuration space, however, the additional long-wavelength power slightly distorts the shape of the BAO peak, and to enable a fair comparison we have added appropriate long-wavelength modes to our theoretical predictions assuming linear theory; specifically, we find that the fitting form $P_{\rm lw}(k) = A\ (k/k_0)^n$, where $A = 3.5 \times 10^4 h^{-3}\ \text{Mpc}^3$, $k_0 = 10^{-3} \kMpc$ and $n = -1.7$, describes well both the long-wavelength excess seen in the power spectrum below $k < 0.01 \kMpc$ and dramatically improves the agreement between the unreconstructed correlation function in theory and {\tt DarkSky}. The contribution to the pre- and post-reconstruction power spectra and correlation function of these long wavelength modes is shown in Figure~\ref{fig:lowk_fit}. Without the long-wavelength correction, the {\tt DarkSky} results do not agree with theory on the large scales to the right of the BAO peak, which should be well-described within linear theory, nor in the BAO ``dip,'' both pre- and post-reconstruction. Due to the ad-hoc nature of our correction, in the remainder of this section we will focus our comparisons on Fourier space, wherein long-wavelength modes must decouple. However, we caution that small, localized features in Fourier space can cause extended distortions in configuration space where data points are highly correlated.  In Figure~\ref{fig:ft_bump}, we show the effect of the $k = 0.1 \kMpc$ bump described in the previous section by additively ``filling'' it with a small, localized Gaussian profile, as shown in the left panel. The effects of this bump, Fourier-transformed, are shown in the right panel: while sub-percent in Fourier space, the $k\simeq  0.1\kMpc$ feature gives rise to visible distortions to the BAO feature in configuration space.

\begin{figure}[h!tb]
    \centering
    \includegraphics[height=\textwidth]{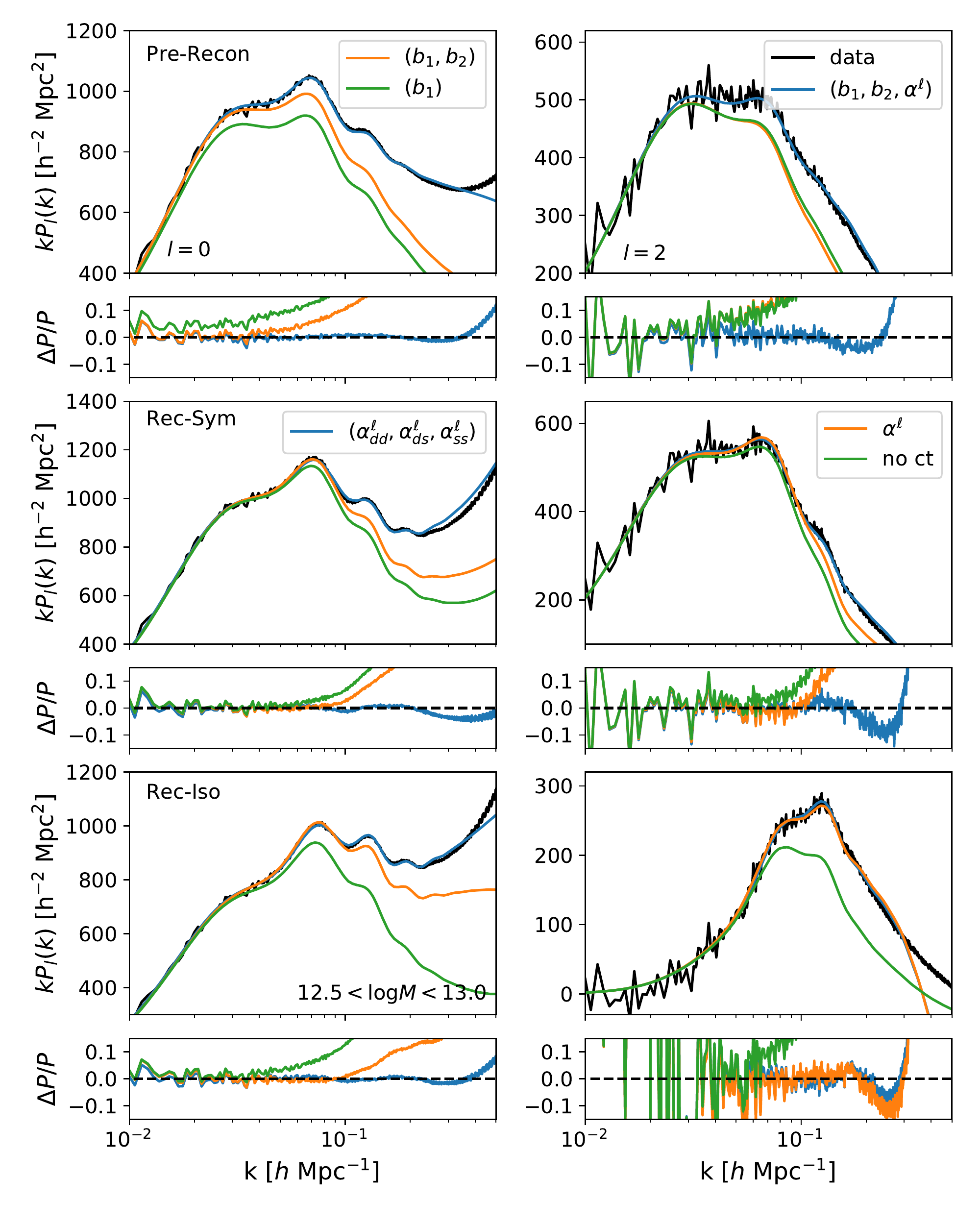}
    \caption{Fits for the pre- and post-reconstruction redshift-space power spectrum monopole (left) and quadrupole (right) for halos in the mass range $12.5 < \log_{10}(M/h^{-1} M_\odot) < 13.0.$ The fractional residuals $\Delta P/P$ are defined in Figure~\ref{fig:real_space_pks}. All spectra were fit using a consistent set of bias parameters $(b_1, b_2) = (0.02, -0.8)$, whose independent contributions are shown in the top row, determined by fitting the pre-reconstruction data, such that only the counterterms were fitted in constructing the curves in the bottom two rows. Our model with the full set of six counterterms---three each for the monopole and quadrupole respectively---fits both the reconstructed monopole and quadrupole in both schemes out to $k = 0.2 \kMpc$ to a few percent and reproduce the phase and amplitude of the oscillatory BAO wiggles.}
    \label{fig:pkrsd_fit}
\end{figure}

\begin{figure}[h!tb]
    \centering
    \includegraphics[height=\textwidth]{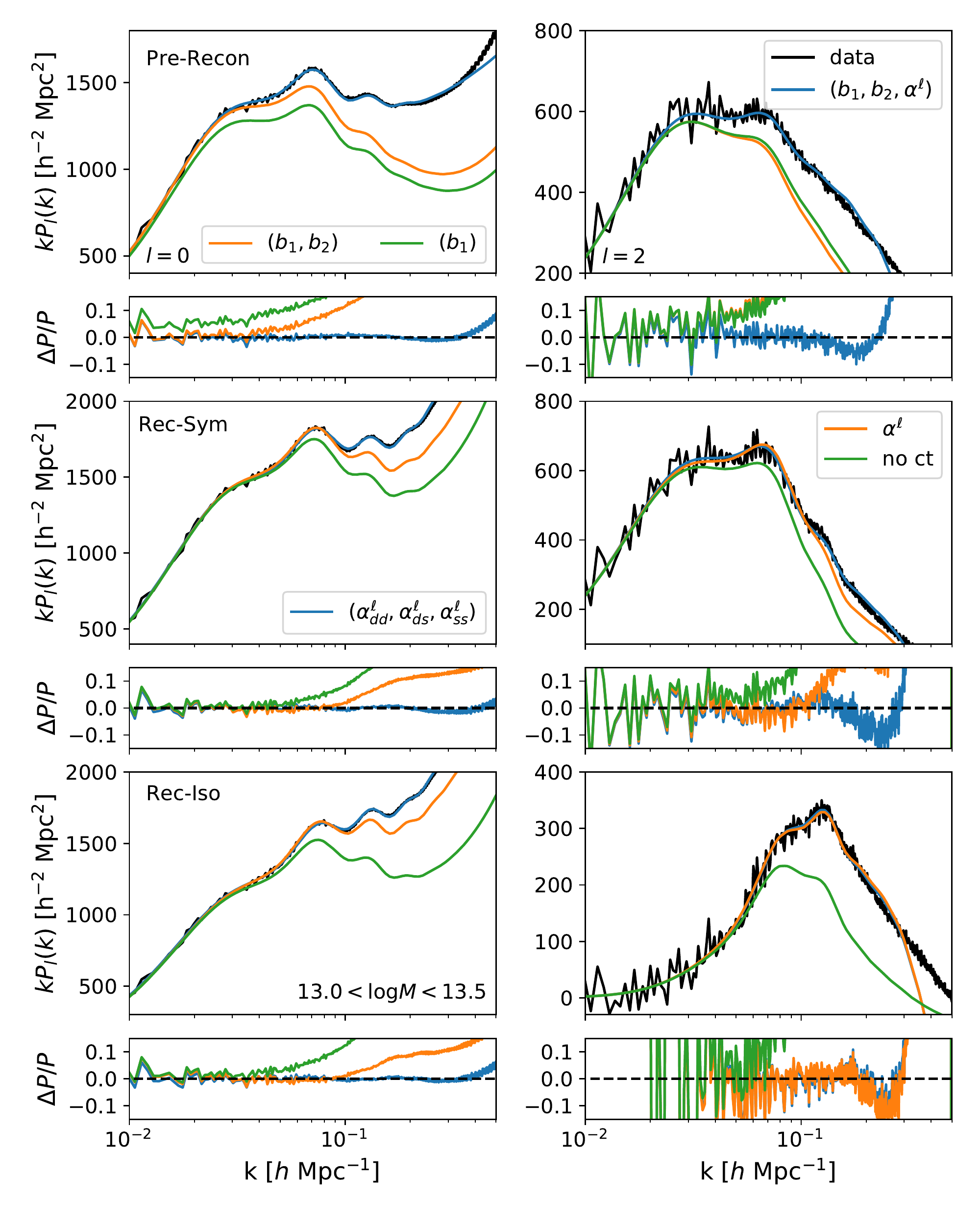}
    \caption{Like Figure~\ref{fig:pkrsd_fit}, but for halos in the mass bin $13.0 < \log(M/h^{-1} M_\odot) < 13.5.$ Here, our model prefers the bias parameters $(b_1, b_2) = (0.23, -1.0)$ and accurately fits the data over a similar range of scales. }
    \label{fig:pkrsd_fit_highmass}
\end{figure}

Finally, fits for the pre- and post-reconstruction power spectra in redshift space are shown in Figure~\ref{fig:pkrsd_fit}.  We have chosen to summarize the angular dependence of the redshift-space power spectrum in terms of its monopole and quadrupole, though our model predicts the full $P(k,\mu)$ and higher multipoles as well. As in real space, we have fitted for the bias parameters $(b_1, b_2)$ using the unreconstructed data and applied the same set of bias parameters to predict the power spectra in both \textbf{Rec-Sym} and \textbf{Rec-Iso}. We adopt the full set of six counterterms, three each  $\alpha^\ell_{dd}, \alpha^\ell_{ds}, \alpha^\ell_{ss}$ for the monopole ($\ell = 0$) and quadrupole ($\ell = 2$), but also explore the possibility of utilizing only one counterterm $\alpha^\ell$ per multipole (corresponding to a derivative bias for both the halo density and velocity). In all cases, our base model with six counterterms fits the data at the percent level or below past $k = 0.2\kMpc$ in both the monopole and quadrupole moments. Notably the Zeldovich approximation produces oscillation-free residuals even in the absence of counterterms (green), with the counterterms providing a physics-based broadband model ($\sim \alpha_{ab}^\ell k^2 P_{\ell,ab}$) that reproduces the N-body results at the percent level. Our fits do not explicitly include nonlinear redshift space distortions such as fingers-of-god, though such effects are perturbatively accounted for by velocity counterterms to lowest order. For completeness, in Figure~\ref{fig:pkrsd_fit_highmass} we show the same fits for the mass bin $13.0 < \log(M/h^{-1} M_\odot) < 13.5,$ where our model fits the data at percent level over a similar range of scales using the parameters $(b_1, b_2) = (0.23, -1.0).$

Lastly, let us comment on the comparison fits in pre- and post- reconstructed cases. 
Given that our shift field, $\chi$, is constructed only from long-wavelength modes explicitly isolated
from observed field, $\delta$, by filtering out the nonlinear scales larger than $k\gtrsim 1/R$, we have no reason to suppose that the perturbative structure of our results will significantly change. In other words, by performing the mapping in Equation~\eqref{eq:rec_map}, we have reconstructed only the long modes, thereby reducing nonlinear smoothing due to large scale (infrared) modes, while the bulk of the small-scale nonlinear modes, as well as FoG effects, should remain unreduced. In addition, Lagrangian perturbation theory (PT) conveniently separates nonlinearities due to long and short modes, exponentially resumming the former while expanding the latter order-by-order \cite{PorSenZal14,VlaWhiAvi15}. Because of this, we do not expect dramatically different PT behavior in the pre- and post-reconstructed results. These arguments are also
supported by Figures~\ref{fig:pkrsd_fit} and \ref{fig:pkrsd_fit_highmass}, which show our model exhibits quantitatively similar degrees of fit pre- and post reconstruction.

\section{Comparison to earlier work}
\label{sec:earlier}

There has been significant theoretical activity in modeling post-reconstruction clustering (see references in the introduction). Our framework encompasses most of these previous perturbation theory expressions when appropriate approximations and phenomenological choices are accounted for. To the best of our knowledge, the framework presented here captures for the first time all of the relevant post-reconstruction effects and is unique in accurately handling both Fourier and configuration space results, in real and redshift space and includes all the bias operators to quadratic order.

\begin{figure}
    \centering
    \includegraphics[width=\textwidth]{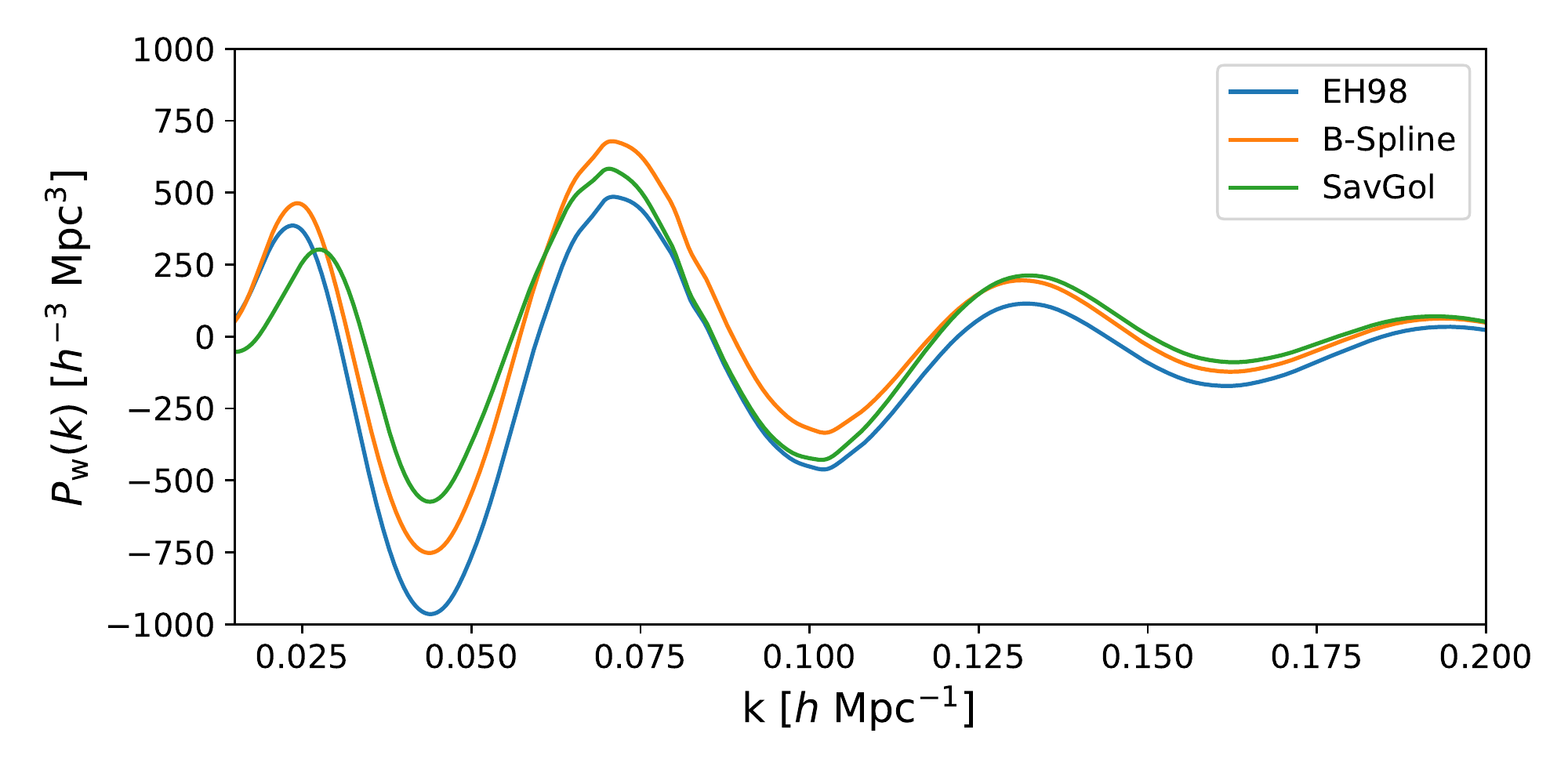}
    \caption{The linear wiggle power spectrum for three choices of $P_{\rm nw}$. The  conventional choice (EH98 \cite{EH1998}) does not accurately capture the large scale power, and we have investigated two possible methods to mitigate this discrepancy: one based on B-splines, described in ref.~\cite{Vlah16} and another based on a Savitsky-Golay filter in $\ln(k)$. The wiggle power spectra isolated using these three methods exhibit visibly different oscillatory behavior.}
    \label{fig:nw_choice}
\end{figure}

Not all models are based on perturbation theory calculations however, and many phenomenological models have been introduced in order to describe the post-reconstruction statistics.  Restricting ourselves just to models of the `standard' reconstruction algorithm \cite{ESSS07}, \S3.1 of ref.~\cite{Xu12} discusses early models (which were of the form $P(k)=B(k)P_{\rm lin}(k)+A(k)$ with $B(k)$ and $A(k)$ smooth functions).  Starting with the first applications to data in ref.~\cite{Xu12} the form used to fit reconstructed power spectra is based upon a split between a ``smooth'' and ``wiggle'' contribution to $P(k) = P_{\rm nw}(k) + \Delta P_{\rm w}(k)$, with a phenomenological damping of the wiggle component motivated by perturbation theory \cite{ESW07}.  In ref.~\cite{Xu12} the parameters of the model were fit to N-body simulations, and this has become common.  This approach has dominated the modeling of observations to date (e.g.~refs.~\cite{Alam17,Carter18,Bautista18} for recent examples) though ref.~\cite{Kazin14} is an example of an analysis that did not take this approach.  However, we note that the choice of the wiggle/no-wiggle split exhibits a certain amount of freedom in the separation of the wiggle and broadband part. This of course implies that, in order to extract accurate information from the e.g.\ BAO, either both wiggle and broadband part have to be modeled to the same level of accuracy, or the extracted wiggle part from the data needs to exactly correspond to the model (see also refs.~\cite{Noda2017, Nishimichi2018} for related discussion). The latter requirement, even though implicitly assumed in most of the current BAO treatments, is rarely subject to performance checks and scrutiny. In this context, it is also worth noting that the common choice of $P_{\rm nw}$ derived in ref.~\cite{EH1998} does not fully capture the broadband linear power spectrum at the precision attained by modern Boltzmann codes. Figure~\ref{fig:nw_choice} shows three possible linear wiggle power spectra, based on no-wiggle spectra computed using the fitting formula from ref.~\cite{EH1998}, B-splines \cite{Vlah16} or a Savitsky-Golay filter; even the latter two, which agree asymptotically with the full linear theory power spectrum, exhibit noticeably different oscillatory behavior. This indicates that extracting the corresponding wiggle spectra from the data is a challenging and sensitive step which can, on the other hand, be avoided if the broadband is included in the theoretical framework. Models phenomenologically relying on a wide separation of scale, assuming scale-independent bias or sufficient smoothness that could be accounted for by nuisance parameters such as $A(k)$ above, might suffer from overall systematic offsets.  Finally, it is also often the case that the nuisance parameters and BAO scaling parameters are not consistent between the configuration-space and Fourier-space analyses (i.e.\ the two do not form a Fourier transform pair) which could prove problematic if fits in both spaces are combined.

By contrast the Zeldovich calculation above gives a consistent framework for understanding the nonlinear smoothing of the BAO feature, both pre- and post-reconstruction, in both configuration and Fourier space. Roughly speaking, the Gaussian smoothing kernel in the empirical model is replaced by a Lagrangian coordinate-dependent kernel $\exp[-k_i k_j A_{ij}(\bq)/2]$.   One might thus hope to formally extract the model for wiggle-only part as an approximation to the calculation presented in the main body of this paper; indeed, such a calculation was performed in ref.~\cite{Vlah16} and extended to terms involving linear bias, redshift space distortions and reconstruction in ref.~\cite{Ding18}\footnote{We note that redshift-space reconstruction model presented in ref.~\cite{Ding18} contains phenomenological damping factors that do not capture the exact behaviour of $X_{ds}$ term given by in Equation \eqref{eqn:Xds}. We repeat this calculation and derive the proper damping factors for the wiggle component in Appendix \ref{app:wiggles}.}. Figure \ref{fig:b1_wiggles} compares the results of our full Zeldovich calculation in the Rec-Sym scheme, with the broadband subtracted out by calculating the corresponding Zeldovich power spectrum using the no-wiggle power spectrum, versus the resummed linear wiggle power spectrum (RWiggle; using the proper exponential damping dependencies given in Appendix \ref{app:wiggles}), for the same linear bias values and with all higher bias terms set to zero. The two are in excellent agreement, especially in the case of the reconstructed power spectrum, with RWiggle slighly underdamping the BAO wiggles towards small scales compared with the full Zeldovich calculation for the unreconstructed power spectrum. 

However, even though RWiggle and the full Zeldovich calculation exhibit a high level of agreement on the shape of the wiggle component, the RWiggle method depends on the separation procedure of the wiggle and broadband components while the full Zeldovich calculation requires no such steps.
Specifically, the Zeldovich calculation deals only with the combination $P^{\rm Zel}_{\rm w} + P^{\rm Zel}_{\rm nw}$, which is obviously invariant under the split, while RWiggle models only the split-dependent $P^{\rm Zel}_{\rm w}$.
This implies that in order to use RWiggle in practical analyses either the broadband part needs to be modeled to equally high accuracy or a highly accurate wiggle extraction procedure is needed in order to guarantee feasible comparison of theoretical model and the data. The latter seems to be a challenging task, potentially subject to systematic offsets and bias. On the other hand, the resulting differences in the wiggle spectrum should still be broadband and could be fit away using nuisance parameters using sufficiently general broadband models.

Finally, our model differs from most in the literature in taking into account higher bias terms such as $b_2$ and $b_s$, allowing us to assess systematic effects introduced by assuming scale-independent bias. These higher biases can contribute both significant broadband power (e.g.\ the top-left panel of Figure~\ref{fig:pkrsd_fit}) and modulate the phase and amplitude of BAO oscillations through mode-coupling effects \cite{PWC09}. However,
explicit calculation shows that the latter effect is only noticeable at very high values of bias. Figure~\ref{fig:wiggles_wb2} shows the effects on the wiggle component of adding nonzero quadratic density and shear biases $b_2$, $b_s$, for bias values $(b_1, b_2) \approx (5,20)$ chosen according to the peak-background split (PBS) on a Press-Schechter mass function \cite{PreSch74}, and assuming $b_s \approx b_2$, as compared to RWiggle. The quadratic density bias, $b_2$, induces an apparent phase shift towards large $k$, and can be seen to be essentially out-of-phase with the linear BAO wiggles; however, these out-of-phase contributions are dramatically reduced by reconstruction. By contrast the shear bias, $b_s$, produces oscillatory features roughly in-phase with the linear theory contributions and is largely unaffected by reconstruction. For completeness, we have also plotted the potential oscillatory contribution of a derivative bias, $b_{\nabla^2}$, which modulates the overall amplitude of the power spectrum and is degenerate with the various counterterms, $\alpha_{ab}$.

To investigate the extent to which the broadband and oscillatory contributions of higher bias terms can be mitigated by a suitable broadband model, we conducted an exploratory ``fit'' of the redshift-space monopole and quadrupole pre- and post-reconstruction in the case where the truth is given by the Zeldovich approximation including nonzero $b_2$ and $b_s$ but fit by an empirical model with only $b_1$, an isotropic BAO scale paramter $\alpha_{\rm BAO}$ and polynomial broadband contributions of the form employed in ref.~\cite{Beu16} before reconstruction. Specifically, we assume an empirical model of the form
\begin{equation}
P_{l,\rm fit}(k) = \alpha_{\rm BAO}^{-3} P_{l,b_1}\Big(\frac{k}{\alpha_{\rm BAO}}\Big) + \frac{a_{1,l}}{k^3} + \frac{a_{2,l}}{k^2} + \frac{a_{3,l}}{k} + a_{4,l} + a_{5,l} k,
\end{equation}
both pre- and post-reconstruction, where $P_{l,b_1}$ denotes redshift-space multipoles in the Zeldovich approximation with all higher biases set to zero. For this exercise we assumed a sample variance limited survey at $z = 0$ and $z = 1.2$ with Gaussian covariances between the monopole and quadrupole, fit up to $k_{\rm max} = 0.25 \kMpc$ and note that the results are independent of survey volume. In Figure~\ref{fig:bao_shift} we have plotted the resulting shifts in the BAO scale assuming PBS values for $b_1$ and $b_2$, taking values for $b_s$ as a function of $b_1$ from ref.~\cite{Abidi18}. At $z = 0$, we find that neglecting higher biases in favor of the empirical model induces shifts of less than half a percent in the BAO scale over a wide range of halo masses both pre- and post-reconstruction, though reconstruction more than halves the forecasted shift for essentially all values of bias surveyed (Figure~\ref{fig:bao_shift}). At $z = 1.2$ the shifts are further reduced, amounting to less than a tenth of a percent across a wide range of bias values prior to reconstruction and essentially vanishing post reconstruction. These shifts would be well-within the margin of error of both current and next-generation surveys like DESI \cite{DESI}, especially post-reconstruction, suggesting that nonlinearities (e.g.\ higher bias) in the power spectrum should not hinder accurate recovery of the BAO signal.  On the other hand, the value of the linear bias, $b_1$, was significantly affected by the choice of broadband model, with fits from the empirical model deviating from the true value by more than five percent in many cases.

\begin{figure}
    \centering
    \includegraphics[width=\textwidth]{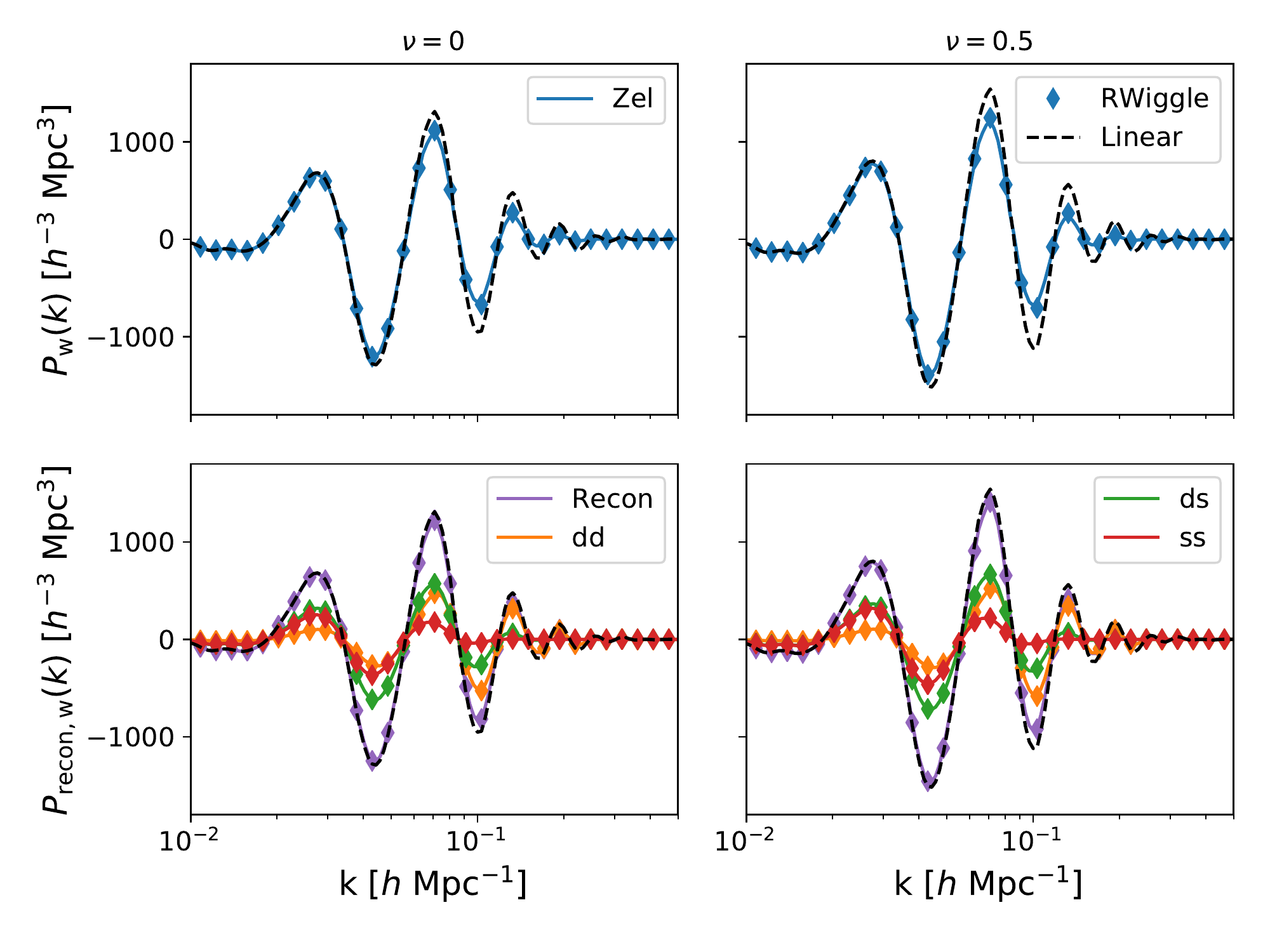}
    \caption{Comparison of Zeldovich with IR-resummed linear theory (RWiggle) for reconstructed and unreconstructed spectra at $z=0$ and $\nu = 0$ and $0.5$ with $b_1 = 0.5$ using Rec-Sym with higher biases set to zero. RWiggle slightly under-predicts damping at high $k$ (but see footnote~\ref{fn:damp_fac}), especially for the unreconstructed power spectra.
    }
    \label{fig:b1_wiggles}
\end{figure}

\begin{figure}
    \centering
    \includegraphics[width=\textwidth]{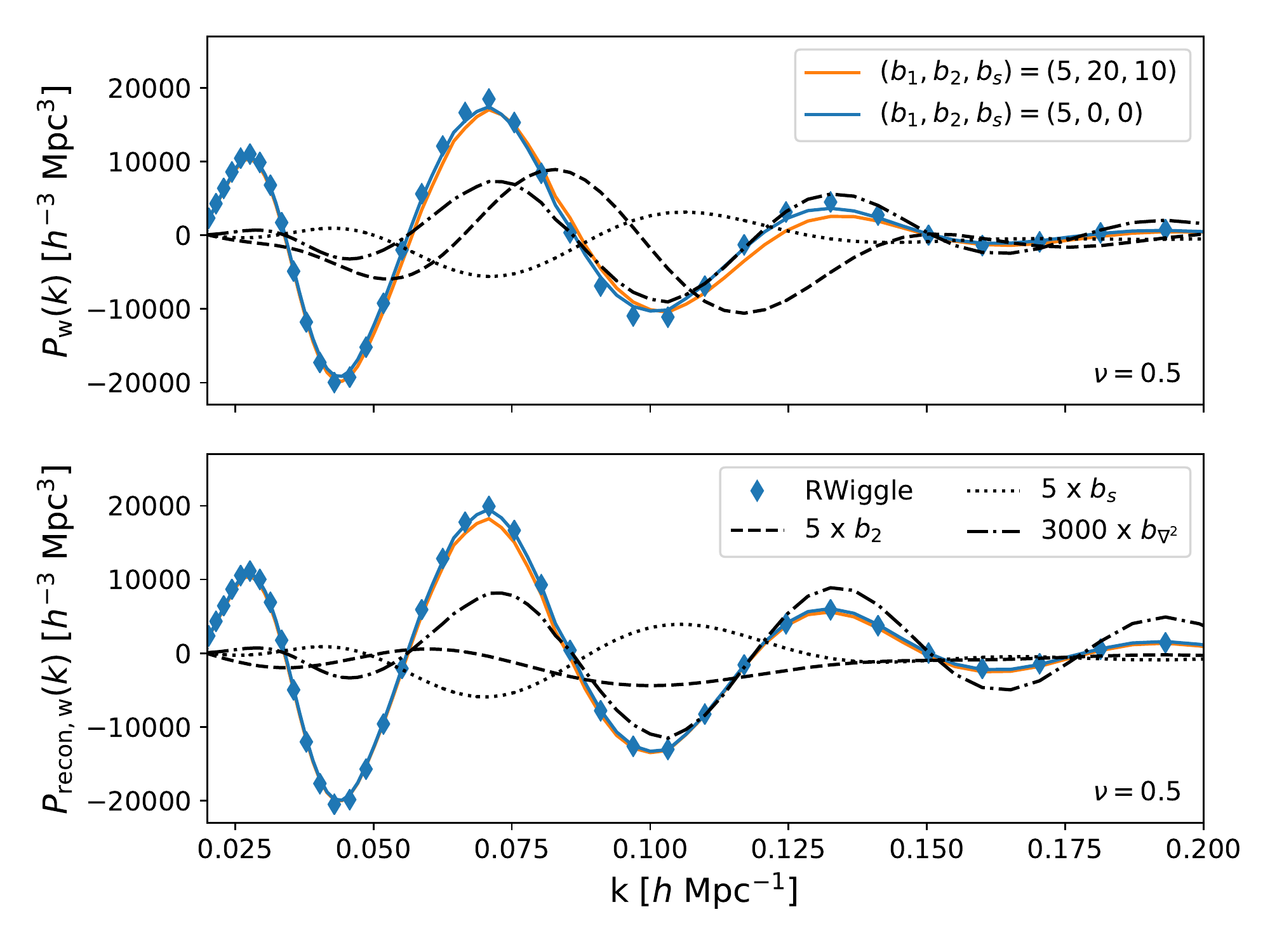}
    \caption{The $z = 0$ Zeldovich power spectrum at $\nu = 0.5$, before and after reconstruction using Rec-Sym, shown with and without contributions from the quadratic bias and shear biases when $(b_1, b_2, b_s) = (5,20,10)$. For comparison, the RWiggle prediction is shown in the diamond points, and the isolated $b_2$ contributions are shown as a black dot-dashed line multiplied by a factor of five. For the unreconstructed spectrum, the $b_2$ contributions (with shear bias set to zero) can be seen to be essentially out-of-phase with the linear theory wiggles and induce a phase shift in the power spectrum. These contributions are greatly reduced in the reconstructed spectrum. The shear contributions, on the other hand, are more-or-less in phase with linear theory and unchanged by reconstruction. For completeness, we have also plotted contributions from a possible derivative bias $b_{\nabla^2}$, which modulate the amplitude of the wiggles in a manner growing with wave number.
    }
    \label{fig:wiggles_wb2}
\end{figure}

\begin{figure}
    \centering
    \includegraphics[width=\textwidth]{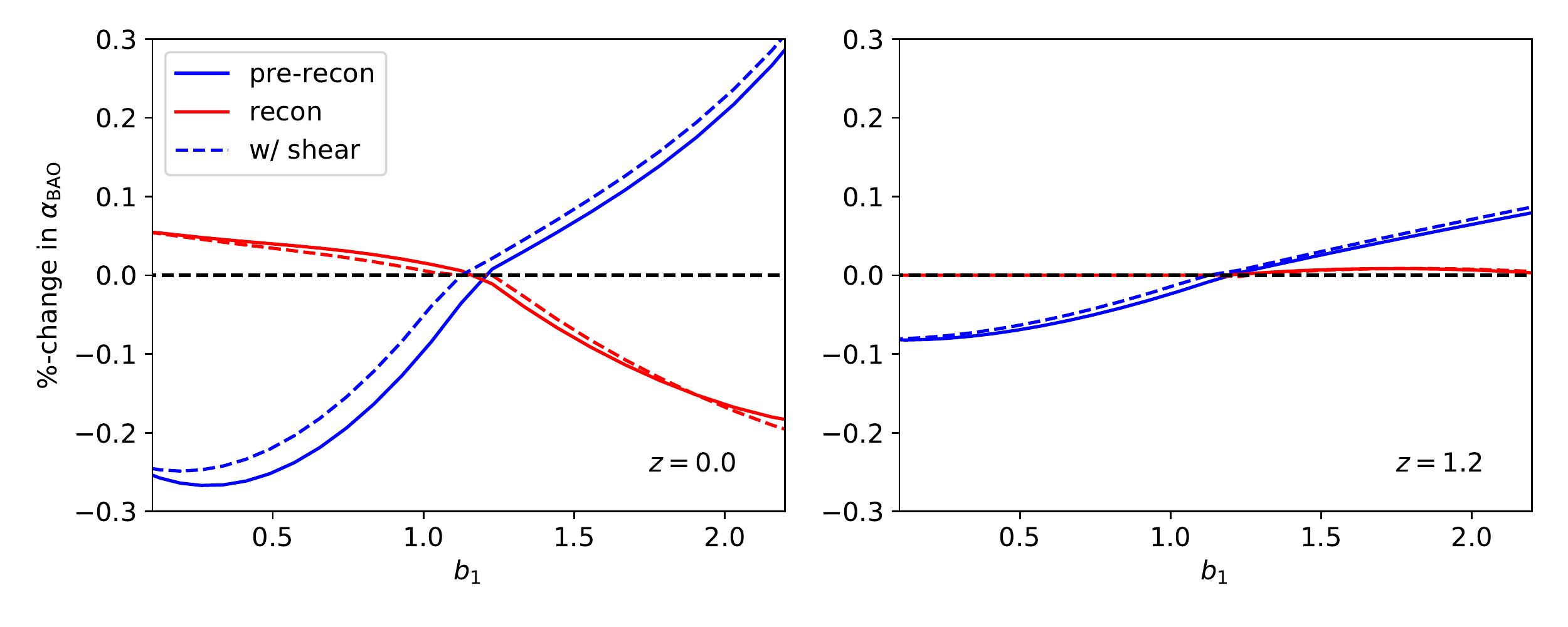}
    \caption{Shifts in the recovered isotropic BAO scale, $\alpha_{\rm BAO}$, in redshift space fit using a model with only $b_1$ nonzero and polynomial broadband contributions in both the monopole and quadrpole, when truth is given by the Zeldovich approximation with nonzero quadratic bias. Values of $b_1$ and $b_2$ were chosen according to the peak-background split, while values for $b_s$ were taken from ref.~\cite{Abidi18}. (Left) Shifts in the BAO scale at $z = 0$. Fitting with the empirical model results in only sub-percent shifts across a wide range of halo masses, which are further more than halved after reconstruction. The solid and dashed lines show the shift with and without the quadratic shear bias $b_s$, whose effect is subdominant to $b_2$. (Right) The same shifts calculated at $z = 1.2$. Even prior to reconstruction, fitting with the empirical model results in less than a tenth of a percent shift in the BAO scale over a wide range of biases; after reconstruction the shift due to nonlinear bias becomes essentially zero.}
    \label{fig:bao_shift} 
\end{figure}

\section{Conclusions}
\label{sec:conclusions}

Baryon acoustic oscillations (BAO) are an important probe of fundamental physics and a prime focus of upcoming surveys such as DESI \cite{DESI} and EUCLID \cite{Euclid}.  The BAO features act as a ``standard ruler'' whose cosmological evolution is largely immune to astrophysical effects but whose signal-to-noise ratio is lowered by nonlinear structure formation.  BAO reconstruction attempts to sharpen the BAO signal by removing some of the nonlinear smearing due to large scale displacements \cite{ESSS07}.  In this paper we develop an analytical model, within the Lagrangian perturbation theory framework, to study the algorithm for density-field reconstruction proposed in ref.~\cite{ESSS07}.  Linear Lagrangian perturbation theory (the Zeldovich approximation) provides an excellent description of these nearly linear displacements and BAO smoothing pre-reconstruction \cite{Zel70,Whi14}, making LPT a promising arena within which to model the effects of reconstruction.

We develop a self-consistent framework with which to calculate the two-point statistics of galaxies, employing a consistent set of parameters to fit the power spectrum and correlation functions, pre- and post-reconstruction in real and redshift space. The broad validity of such LPT models allows for joint fits to the pre- and post-reconstruction two-point statistics enabling e.g.~a fit for redshift-space distortions and the linear growth rate, $f\sigma_8$, simultaneously with Alcock-Paczynski distortions constrained by BAO analyses \cite{Whi15}. Based on ref.~\cite{VlahWhi18}, we derive explicit formulae, to calculate the redshift-space power spectrum within the Zeldovich approximation, both pre- and post-reconstruction, as an infinite series of spherical Bessel transforms. Our model updates the developments for the reconstructed correlation function in ref.~\cite{Whi15}, and is -- as far as we are aware -- the first model of reconstruction to include a consistent set of bias terms up to quadratic order, including shear and derivative biases. We show that the oscillatory behavior induced by the quadratic density bias, $b_2$, are out of phase with the linear BAO feature and greatly reduced post-reconstruction, while those due to the quadratic shear bias, $b_s$, are in-phase and essentially unchanged. In addition, we show that each multipole moment of the reconstructed power spectrum should be, to lowest order, corrected for by a set of three counterterms each, which perturbatively correct both nonlinear smoothing and broadband power.

We compare our analytic predictions with N-body data from the {\tt DarkSky} simulation \cite{skillman14} at $z = 0$, focusing on halos between $12.5 < \log_{10}(M/M_\odot) < 13.0$. Our base model, involving only $b_1$ and $b_2$ and appropriate counterterms, jointly fits the pre-reconstruction real-space power spectrum and redshift-space monopole out to $k = 0.4 \kMpc,$ and the quadrupole out to $k = 0.2 \kMpc,$ reproducing the oscillatory BAO wiggles in the data with high fidelity. Our model with the same bias parameters performs equally well in configuration space around the BAO scale, though we found it necessary to correct for a large excess in large-scale power encountered in the {\tt DarkSky} data. Utilizing the same values for the bias parameters but allowing counterterms to vary, we find that our model performs similarly in real space post-reconstruction for smoothing scales $R = 15$ and $20\, h^{-1}$Mpc, reproducing both the oscillatory features and broadband past $k = 0.2\kMpc$, but fails to reproduce the oscillatory features when $R=10\,h^{-1}$Mpc, likely due to the fact that we have worked to lowest order and at $z = 0$ displacements on that scale are significantly nonlinear. We point out a less severe feature in the residuals at $k = 0.1\kMpc$ that diminishes with larger smoothing scales which we believe arise from higher order terms and caution that neither our calculation nor the standard reconstruction algorithm take these into account.  A more complete, iterative reconstruction scheme (e.g.\ ref.~\cite{Schmittfull17}) may reduce these features. The modeling of these nonlinearities, and possible remedies, are beyond the scope of this paper, but as an exploratory example we calculate the effects of one possible nonlinearity due to the mapping between Eulerian and Lagrangian coordinates in Appendix~\ref{app:EulerianPosition}.

Our model also predicts the multipole moments of the redshift-space power spectrum and correlation functions in both of the redshift-space schemes (\textbf{Rec-Sym} and \textbf{Rec-Iso}) we consider.  This is critical in order for it to be applied to data, since the most constraining BAO measurements are performed in redshift space.  The model provides a good fit to the monopole and quadruople moments of $P(k)$ measured in {\tt DarkSky} in both the \textbf{Rec-Sym} and \textbf{Rec-Iso} schemes for smoothing scales of $R=15\,h^{-1}$Mpc or larger (at $z=0$).  Again, for smaller smoothing scales the Zeldovich model differs from the N-body results (as expected).  These effects would be smaller at higher redshift, where the theory is more likely to be applied.

Finally, there exists an extensive literature studying the modeling of reconstruction and the BAO signal, and we compare our model to several existing alternatives. One popular technique, based on ref.~\cite{ESW07}, is to separate the power spectrum into a smooth ``no wiggle'' component and an oscillatory ``wiggle component,'' and to damp the latter by an exponential factor fit to simulations while supplementing the former with a polynomial in wavenumber to fit the broadband power. This technique can be more rigorously derived as a particular resummation of the nonlinear contribution of long-wavelength modes much like our Zeldovich calculation itself \cite{Vlah16,Ding18}, in which case the damping parameters can be derived theoretically. When the ``wiggle'' components are isolated we find that the latter is in excellent agreement with our Zeldovich calculation, particularly after reconstruction. In Appendix~\ref{app:wiggles} we re-derive the IR-resummed ``wiggle'' power spectrum (RWiggle) directly within our Zeldovich framework, updating the exponential damping for the cross term $P^{ds}$. We highlight that our Zeldovich framework naturally encompasses broadband effects, while methods depending on wiggle/no-wiggle splitting might be subject to additional systematic offsets and biases. These could originate from the fact that the wiggle/no-wiggle splitting is not unique, and thus relies on correctly predicting the broadband or extracting the corresponding wiggle part from the data to high accuracy.  On the other hand, the Zeldovich framework correctly captures broadband power over a large range of scales in addition to reproducing the oscillatory features in the reconstructed power spectrum.  In fits to N-body data, we show how counterterms correct the sharpness of the BAO feature and broadband power simultaneously and consistently. Moreover, our model goes beyond linear bias to include quadratic density and shear bias, which we show contribute oscillatory terms to $P(k)$ that vary independently in amplitude and phase.

We close by noting a few avenues for future work. An obvious extension of our model is to include nonlinearities arising both from gravitational clustering and the reconstruction itself (e.g.\ Appendix~\ref{app:EulerianPosition}). The former may be most easily included in the context streaming models \cite{VlaCasWhi16,VlahWhi18}, wherein the real-space modifications due to reconstruction and those proportional to the growth rate $f$ can be separately treated as modifications to the statistics of the galaxy density and galaxy density-weighted velocities, respectively, and which in addition have the advantage of resumming biased contributions to redshift-space distortions as well as nonlinear redshift-space phenomena like fingers-of-god. It is, however, not a-priori obvious which type of nonlinearity will present the most significant corrections. Other fruitful avenues would be to investigate the impact of wrong parameters on reconstruction (e.g.\ refs.~\cite{Sherwin19,Carter19}) or to update the present treatment to newer reconstruction techniques. Finally, one could investigate the utility of our model for upcoming surveys like DESI \cite{DESI} or Euclid \cite{Euclid}. These surveys will operate at higher redshifts where our calculations should perform even better, and our model will be a natural arena in which to understand the effects of highly biased tracers and the effects of cosmic evolution (e.g.\ evolving $b$ and $\sigma_8$) on the BAO feature measured in broad redshift bins.

We have publicly released our codes for configuration\footnote{https://github.com/martinjameswhite/ZeldovichRecon} and Fourier\footnote{https://github.com/sfschen/ZeldovichReconPk} space reconstruction, with the hope that they will be useful to other researchers.  We have checked that the Hankel transform of the Fourier space code agrees, term by term, with the configuration space code to better than 1\%, except very close to zero crossings.

\section*{Acknowledgments}
We thank Hee-Jong Seo and Florian Beutler for useful discussions.
SC is supported by the National Science Foundation Graduate Research Fellowship (Grant No. DGE 1106400) and by the UC Berkeley Theoretical Astrophysics Center Astronomy and Astrophysics Graduate Fellowship.
M.W.~is supported by the U.S.~Department of Energy and by NSF grant number 1713791.
This research used resources of the National Energy Research Scientific Computing Center (NERSC), a U.S.\ Department of Energy Office of Science User Facility operated under Contract No.\ DE-AC02-05CH11231.
This work made extensive use of the NASA Astrophysics Data System and of the {\tt astro-ph} preprint archive at {\tt arXiv.org}.

\appendix

\section{Cross-spectra correlators}
\label{sec:correlators}

In this appendix we give analytic expressions for the two-point functions required to calculate cross-spectra, which are slightly different from those required to calculate the auto-spectra more commonly seen in the literature. 

The two-point function for the Lagrangian displacement between two species separated by Lagrangian distance $\bq$ is given by
\begin{equation}
    A^{ab}_{ij}(\bq) = \langle \bPsi^a_i \bPsi^a_j \rangle + \langle \bPsi^b_i \bPsi^b_j \rangle - 2 \langle \bPsi^a_i(\bq_2) \bPsi^a_j(\bq_1) \rangle \equiv X^{ab}(q)\ \delta_{ij} + Y^{ab}(q)\ \hat{q}_i \hat{q}_j
\end{equation}
where
\begin{align}
    X^{ab}(q) &= \frac{2}{3} \int \frac{dk}{2\pi^2} \Big[ \ \frac{1}{2} \Big( P^{aa}_L(k) + P^{bb}_L(k) \Big) - \Big( j_0(kq) + j_2(kq) \Big) P^{ab}_L(k) \Big]  \nonumber \\
    Y^{ab}(q) &= 2 \int \frac{dk}{2\pi^2} \ j_2(kq) P^{ab}_L(k).
    \label{eqn:XYdef}
\end{align}
Note that for cross spectra $X^{ab}(q)$ does not in general vanish as $q \rightarrow 0$. Similarly we have
\begin{equation}
    U^b_i = \langle \Delta^{ab}_i \delta_0(\bq_2) \rangle \equiv U^b(q) \hat{q}_i, \quad U^a_i = \langle \Delta^{ab}_i \delta_0(\bq_1) \rangle \equiv U^a(q) \hat{q}_i
\end{equation}
where
\begin{equation}
    U^{a}(q) = - \int \frac{dk\ k}{2\pi^2}\ j_1(kq) P^{am}_L(k)
\end{equation}
and $P^{am}$ is the linear theory cross spectrum between tracer $a$ and matter, and the corresponding expression for $U^b$ follows by direct substitution.

Finally, the non-scalar shear correlators are given by
\begin{equation}
    V_i^{ab} = V^a(q) \hq_i, \quad \Upsilon^a_{ij} = X^a_{s^2}(q) \delta_{ij} + Y^a_{s^2}(q) \hq_i \hq_j
\end{equation}
where the functions of $q$ are given by
\begin{equation}
    V^a(q) = 2\int \frac{dk\ k}{2\pi^2}\ P^{am}_L(k)\ \Big[ \frac{4}{15}j_1(kq) - \frac{2}{5}j_3(kq) \Big]\ \int \frac{dk\ k^2}{2\pi^2}\ P^{mm}_L(k)\ j_2(kq)
\end{equation}
and
\begin{align}
    X_{s^2}(q) = 4 (\mathcal{J}_3^{a})^2,\quad Y_{s^2}(q) = 6 (\mathcal{J}_2^a)^2 + 8 \mathcal{J}_2^a \mathcal{J}_3^a + 4 \mathcal{J}_2^a \mathcal{J}_4^a + 4 (\mathcal{J}_3^a)^2 + 8 \mathcal{J}_3^a \mathcal{J}_4^a + 2 (\mathcal{J}_4^a)^2
\end{align}
where following refs.~\cite{Whi14,VlaCasWhi16} we have defined
\begin{align}
    \mathcal{J}_2^a &= \int \frac{dk\ k}{2\pi^2}\ P^{am}_L(k)\ \Big[\frac{2}{15}j_1(kq) - \frac{1}{5}j_3(kq) \Big] \\
    \mathcal{J}_3^a &= \int \frac{dk\ k}{2\pi^2}\ P^{am}_L(k)\ \Big[-\frac{1}{5}j_1(kq) - \frac{1}{5}j_3(kq) \Big]\\
    \mathcal{J}_4^a &= \int \frac{dk\ k}{2\pi^2}\ P^{am}_L(k)\ j_3(kq).
\end{align}
The remaining scalar shear correlators, $\zeta$ and $\chi^{12}$, are identical to those found in evaluating the auto-spectrum, and we refer readers to refs.~\cite{Whi14,VlaCasWhi16}.

\section{The pre- and post-reconstruction Zeldovich propagator}
\label{app:zel_prop}

In this appendix we give expressions for the normalized cross-spectrum between the initial and final or reconstructed field.  This is essentially a correlation coefficient, though it is also referred to as the propagator \cite{Crocce08}.  Specifically we define $G_a(\bk) = \langle \delta_0(-\bk) \delta_a(\bk) \rangle/ \langle \delta_0(-\bk) \delta_0(\bk) \rangle$, within the Zeldovich approximation, which quantifies the extent to which a tracer field $a$ is (de)correlated with the initial density $\delta_0$, and apply our results to derive the reconstructed field. Our results generalize those in ref.~\cite{Noh09} to include halo bias.

As defined, the propagator $G_a$ is a special case of the cross spectrum and can be evaluated using Equation~\ref{eqn:cross_spectra} by assuming that the linear field $\delta_0$ is a tracer $b$ with displacement $\bPsi^b = 0$ and bias functional $F^b = \delta_0$, such that any Lagrangian two-point functions involving the displacement $\bPsi^b$ (e.g. $U^b$) or higher biases (e.g. $b_2^b$) vanish identically. Unlike in the conventional case, however, $F^b$ does not have a zero order piece equal to unity--- we can thus compute our result directly by taking the derivative of Equation~\ref{eqn:cross_spectra} with respect to $b_1^b$ with the above assumptions. This gives
\begin{equation}
    P_L(k)G_a(\bk) = \dtq e^{i \bk \cdot \bq}\ e^{-k^2\Sigma_{aa}^2/4} \Big[i k_i U^a_i + b_1^a \xi_L \Big] = e^{-k^2\Sigma_{aa}^2/4} \Big(P^{am}(k) + b_1^a P^{mm}(k) \Big),
\end{equation}
where we have used that
\begin{equation}
    A^{a0}_{ij}(\bq) = \langle \bPsi^a_i \bPsi^a_j \rangle \equiv \frac{1}{2} \Sigma_{aa}^2 \delta_{ij},
\end{equation}
only receives ``half'' of the zero-point contribution c.f.~the power spectrum (where $\langle\bPsi^b\bPsi^b\rangle\neq 0$). Note that all higher bias contributions vanish. The generalization to the redshift space field can be straightforwardly accomplished by multiplying by appropriate factors of $R_{ij}$ in the numerator, though we will focus on real space in this appendix as RSD introduce an equally important but parallel form of decorrelation into the problem.

From the above results, the reconstructed-field ($\delta_{\rm rec} = \delta_d - \delta_s$) propagator can be written as $G_{\rm rec} = G_{d} - G_{s}$, where
\begin{equation}
    G_d(\bk) = \frac{e^{-k^2\Sigma_{dd}^2/4} \Big(P^{dm}(k) + b_1 P^{mm}(k) \Big)}{P_L(k)}
    \quad , \quad
    G_s(\bk) = \frac{e^{-k^2\Sigma_{ss}^2/4} P^{sm}(k)}{P_{L}(k)},
\end{equation}
where the various linear spectra are defined as in Equation~\ref{eqn:linspec}. The real-space post-reconstruction propagator is then
\begin{equation}
    G_{\rm rec}(\bk) = e^{-k^2\Sigma_{dd}^2/4} \left[ (1 - \mathcal{S}(k) + b_1 \right] + \mathcal{S}(k) e^{-k^2\Sigma_{ss}^2/4}.
\end{equation}
The expression for $G_{\rm rec}$ helps to quantify how much of the decorrelation between the initial conditions and the final field arises due to bulk motions, and the manner in which this can be restored by the standard reconstruction algorithm. Roughly speaking, reconstruction reduces the decorrelation from the full matter $\Sigma^2$ to $\Sigma^2_{dd}$ past the smoothing scale for the matter piece, with the correlation at low $k$ close to unity assuming that the damping due to $\Sigma_{ss}^2$ there is negligible.

\section{Integrals for redshift space distortions via direct Lagrangian expansion}
\label{sec:rsd_integrals}

In this section we describe how to perform the three-dimensional integrals that occur when calculating redshift space power spectra in the Lagrangian formalism.

\subsection{Angular Integrals}
\label{sec:mu_ints}

Calculations in Lagrangian perturbation theory frequently require evaluating integrals of the form
\begin{equation}
    I_{\ell}^{(n)}(A,B) = \frac{1}{2} \int^1_{-1} d\mu\ \mu^{2\ell+n} e^{iA\mu + B\mu^2}.
\end{equation}
These integrals can be conveniently expressed as infinite sums of spherical Bessel functions \cite{VlaSelBal15,VlahWhi18}, e.g.
\begin{align}
    I_\ell^{(0)} &= \frac{(-1)^\ell e^B}{B^\ell} \sum_{n=0}^{\infty}\ U(-\ell,n-\ell+1,-B)\ \Big(\frac{-2B}{A} \Big)^n j_n(A) \nonumber \\
    I_\ell^{(1)} &= i \frac{(-1)^\ell e^B}{B^\ell} \sum_{n=0}^{\infty}\ U(-\ell,n-\ell+1,-B)\ \Big(\frac{-2B}{A} \Big)^n j_{n+1}(A) \nonumber \\
    I_\ell^{(2)} &= \frac{(-1)^\ell e^B}{B^\ell} \sum_{n=0}^{\infty}\ \Bigg[ U(-\ell,n-\ell+1,-B) + \frac{n}{B}\ U(-\ell,n-\ell,-B) \Bigg]\ \Big(\frac{-2B}{A} \Big)^n j_{n}(A),
    \label{eqn:mu_ints}
\end{align}
where $U(a,b,z)$ denotes the confluent hypergeometric function of the second kind.

\subsection{Direct Lagrangian Expansion: M{\sc i}}
\label{sec:MI}

Calculating the power spectrum in redshift space within the Zeldovich approximation requires a few extra steps when compared to the calculation in real space due to the line-of-sight dependence of RSD. In the following two sections we extend the two methods presented in ref.~\cite{VlahWhi18}, M{\sc i} and M{\sc ii}, to include bias terms up to quadratic order. These two methods correspond, roughly speaking, to active and passive transformations in Fourier space via $R_{ij} = \delta_{ij} + f \hat{n}_i \hat{n}_j$, respectively. For the general case, we found M{\sc ii} to be somewhat more convenient, and will therefore offer only a cursory description of M{\sc i}, except for a special case involving the displaced-shifted field cross spectrum in \textbf{Rec-Iso}.

In M{\sc i}, the (halo auto-) power spectrum in redshift space is given by
\begin{equation}
    P_s(k) = \int d^3q\ e^{i \bk \cdot \bq - \frac{1}{2} k_i k_j A^{s}_{ij}} \Bigg[1 + 2ib_1 k_i U^{s}_i + b_1^2 \xi_L + ... \Bigg],
\end{equation}
where the superscript $s$ denotes tensor quantities transformed by $R_{ij}$, e.g. $U^{s}_i = R_{ij} U_j$. Each quantity in the above integral can be written in terms of scalar functions and dot products between three unit vectors ($\hq$, $\hat{k}$ and $\hat{n}$) whose angular structure underlies redshift space distortions; these are, in particular,
\begin{equation}
    \hat{n} \cdot \hat{k} = \nu, \quad \hat{q} \cdot \hat{k} = \mu, \quad \hat{q} \cdot \hat{n} = \mu \nu + \sqrt{1-\mu^2}\,\sqrt{1-\nu^2}\,\cos\phi,
\end{equation}
where $\phi$ is the azimuthal angle in a polar coordinate system where the zenith is given by $\hat{k}$ and the plane $\phi = 0$ is spanned by $\hat{k}$ and $\hq$.
The effect of the transformation $\mathbf{R}$ can then be captured by how the usual tensor basis $\hq_i, \delta_{ij},$ etc. is affected:
\begin{align}
    k_i k_j \delta_{ij} &\rightarrow k_i k_j R_{in} R_{jn} = k^2 \left[1 + f(2+f) \nu^2\right] \\
    k_i \hq_i &\rightarrow k_i R_{ij} \hq_j = k \mu \left[1 + f\nu^2 + f\nu^2 \gamma(\mu,\nu) \cos\phi\right] \quad ,
\end{align}
where we have defined $\gamma(\mu,\nu) = \sqrt{1-\mu^2}\,\sqrt{1-\nu^2}/\mu\nu.$ The azimuthal dependence of $A^{s}_{ij}$ requires us to calculate polar-coordinate integrals of the form \cite{VlahWhi18}
\begin{equation}
    I_n(f, \mu,\nu) = \int^{2\pi}_0 \frac{d\phi}{2\pi}\ e^{-\frac{1}{2}k^2 Y (\alpha_2 \gamma \cos\phi + \alpha_3 \gamma^2 \cos^2\phi)\mu^2} (\gamma \cos\phi)^n
\end{equation}
where $\alpha_2 = 2 f \nu^2 (1 + f\nu^2)$ and $\alpha_3 = f^2\nu^4$. These can be calculated as analytic power series in $\mu$ by taking derivatives of the identity
\begin{equation}
I_\phi \left( \alpha, \beta, \mu \right) = \int_0^{2\pi} \frac{d\phi}{2\pi}~e^{ \alpha \mu \sqrt{1-\mu^2} \cos\phi
+ \beta (1 - \mu^2) \cos^2\phi }
=  \sum_{\ell=0}^\infty F_{\ell}(\alpha, \beta) \left(\alpha^2 \mu^2/\beta \right)^{\ell}~
\label{eqn:mi_phi_integral}
\end{equation}
where
\begin{align}
F_{\ell}(\alpha, \beta) &= \sum_{m=0}^\ell
\frac{\Gamma(m+\frac{1}{2})}
     {\pi^{1/2}\Gamma(m+1)\Gamma(1+2m-\ell)\Gamma(2\ell-2m+1)}
     \left( - \frac{\beta^2}{\alpha^2} \right)^m \nonumber\\
&\hspace{2.4cm} \times  M\left(\ell-2m; \ell-m+\tfrac{1}{2};  \frac{\alpha^2}{4\beta} \right)
 M\left(m+\frac{1}{2};m+1; \beta \right) \nonumber
\end{align}
and $M(a,b,z)$ are hypergeometric functions of the first kind. Each term proportional to $\mu^n$ can in turn be evaluated as a spherical Bessel transform as in Equation \ref{eqn:mu_ints}.

A simplified but demonstrative example of this calculation can be found in the calculation of the displaced-shifted cross spectrum in redshift space reconstruction via \textbf{Rec-Iso}. The $H^{0}_\ell$ expansion in Equation~\ref{eqn:methodii_cross} is essentially Equation~\ref{eqn:mi_phi_integral} in the limit where $\beta \rightarrow 0.$ To proceed from Equation~\ref{eqn:methodii_cross}, we can use the identities in \ref{sec:mu_ints} and refactor the resulting double sum over $n$ and $\ell$ to get
\begin{align}
    P^{(ds)}(\bk) &=  e^{-\frac{1}{2}k^2 (\alpha_{0} \Sigma^{(dd)^2} + \Sigma^{(ss)^2})}\ 4\pi \sum_{n=0}^{\infty} \int dq\ q^2 e^{k^2 (1 + f\nu^2)(\tilde{X}^{ds}+\tilde{Y}^{ds})} \Big(\frac{-2k\tilde{Y}^{ds}}{q} \Big)^n\ (1+f\nu^2)^n \nonumber \\
    &\left[K^{(0)}_n(q) j_n(kq) -  b_1 k U^{d}(q) K^{(0)}_n(q) j_{n+1}(kq) - \frac{1}{2} b_2 k^2 U^{d}(q)^2 K^{(2)}_n(q) j_n(kq) + \cdots  \right]
\end{align}
where the redshift-space kernels are given by 
\begin{align}
    K^{(0)}_n(q) &= \sum_{\ell=0}^{\infty} \Big(-\frac{f\nu\sqrt{1-\nu^2}}{1+f\nu^2} \Big)^\ell H_\ell(A)\ U(-\ell,n-\ell+1,-B) \nonumber \\
    K^{(2)}_n(q) &= \sum_{\ell=0}^{\infty} \Big(-\frac{f\nu\sqrt{1-\nu^2}}{1+f\nu^2} \Big)^\ell H_\ell(A)\ \left[ U(-\ell,n-\ell+1,-B)+\frac{n}{B}\ U(-\ell,n-\ell,-B)\right] \nonumber
\end{align}
and, as before, $A = k^2f\nu\sqrt{1-\nu^2}\,\tilde{Y}^{ds}$ and $B = k^2(1+f\nu^2)\,\tilde{Y}^{ds}$. Deriving these kernels for the other terms is entirely analagous\footnote{The mater contribution was given in ref.~\cite{VlahWhi18}.}.

\subsection{Direct Lagrangian Expansion: MII}
\label{sec:MII}

An equivalent approach to rotating each Lagrangian displacement $\bPsi$ by $\vb{R}$ is to instead passively transform the Fourier basis by $\vb{R}^{\rm T}$, such that the wavenumber is given by $K_i \equiv R_{ij} k_j$. Defining $\mu = \hat{K} \cdot \hat{q}$ as the angle between the transformed wave vector and Lagrangian separation $\bq$, we have the dot products
\begin{align}
    &K \cdot K = k^2\left(1 + f(2+f)\nu^2\right) \nonumber \\
    &\hat{K} \cdot \hat{n} = k \nu (1+f) \nonumber \\
    &\hat{n} \cdot \hat{q} = \frac{\mu \nu (1+f)}{\sqrt{1+f(2+f)\nu^2}} + \sqrt{1-\mu^2} \sqrt{1 - \Big( \frac{\nu^2(1+f)^2}{1+f(2+f)\nu^2}\Big)} \,\cos\phi \nonumber \\
    &k \cdot q = kq \Big(c\mu - s\sqrt{1-\mu^2}\cos\phi \Big), \; c = \frac{1+f\nu^2}{\sqrt{1+f(2+f)\nu^2}}, \; s = \frac{f\nu\sqrt{1-\nu^2}}{\sqrt{1+f(2+f)\nu^2}}. 
    \label{eqn:MII_dots}
\end{align}
The power spectrum in this frame is then simply given by substituting $K$ for $k$ in the real space expression; e.g.\ the term in the exponential can be written in terms of the vector $\textbf{K}$ as
\begin{equation}
    -\frac{1}{2} k_i k_j A^s_{ij} = -\frac{1}{2} k_i k_j R_{il} R_{jm} A_{lm} = -\frac{1}{2} K_\ell K_m A_{\ell m}.
\end{equation}
In redshift space, unlike in real space, the Fourier product $\bk \cdot \bq$ requires azimuthal angle integrals due to the appearance of $\cos\phi$ in the final line of \ref{eqn:MII_dots}; however, in M{\sc ii} the azimuthal dependence is factored entirely into the Fourier factor $i \bk \cdot \bq$ such that the $\phi$ integral can be performed analytically:
\begin{align}
     P_s(k) &= \int dq\ d\mu\ q^2\ e^{ikqc\mu- \frac{1}{2}K^2(X+Y\mu^2)}\ \Bigg(\int d\phi\ e^{-ikqs\sqrt{1-\mu^2} \cos\phi} \Bigg)\ \Big[1 +  2i b_1 K_i U_i(\bq) + ...\Big]  \nonumber \\
     &= 2\pi \int dq\ d\mu\ q^2\ e^{ikqc\mu- \frac{1}{2}K^2(X+Y\mu^2)}\ J_0(kqs\sqrt{1-\mu^2})\ \Big[1 +  2i\mu\ b_1 K U(q) + ...\Big]. 
\end{align}
The remaining $q$ and $\mu$ integrals can then be calculated using the usual combination of spherical-Bessel decompositions and Hankel transforms with the help of the identity \cite{VlahWhi18}:
\begin{equation}
    \int d\mu\ e^{i\mu A + \mu^2 B} J_0(C\sqrt{1-\mu^2}) = 2e^B\ \sum_{\ell=0}^{\infty}  \Big(\frac{-2}{\rho} \Big)^m  \tilde{G}^{(0)}_\ell(A,B,\rho) j_\ell(\rho) \quad ,
\end{equation}
where $\rho = \sqrt{A^2+C^2}$ and the function $\tilde{G}_m(A,B,\rho)$ is given by
\begin{equation}
    \tilde{G}^{(0)}_m(A,B,\rho) = \sum_{n=m}^{\infty} f_{nm} \Big(\frac{BA^2}{\rho^2} \Big)^n  {}_2F_1(\frac{1}{2}-n,-n; \frac{1}{2}-m-n; \frac{\rho^2}{A^2}) \nonumber,
\end{equation}
$_2F_1$ is the ordinary hypergeometric function, and we have defined
\begin{equation}
    f_{nm} = \frac{\Gamma(m+n+\frac{1}{2})}{\Gamma(m+1)\Gamma(n+\frac{1}{2})\Gamma(1-m+n)}
    \quad .
\end{equation}
In our specific case, $\rho = kq$, $B = - K^2 Y/2$ and $A = kqc.$ Defining
\begin{equation}
    I_n = \int d\mu\ (i\mu)^n e^{iA\mu + \mu^2 B} J_0(C\sqrt{1-\mu^2}) \equiv 2e^B \sum_{l=0}^{\infty} \Big(\frac{-2}{\rho}\Big)^l \tilde{G}^{(n)}_l(A,B,\rho) j_\ell(\rho)
\end{equation}
such that $I_n = I_0^{(n)}(A)$ is the n$^{\rm th}$ full derivative of $I_0$ with respect to $A$, we have recursively
\begin{equation}
    \tilde{G}^{(n)}_{l} =  \frac{d\tilde{G}^{(n-1)}_l}{dA} + \frac{A}{2}\ \tilde{G}^{(n-1)}_{l-1} .
\end{equation}
The first two derivatives of $\tilde{G}^{(0)}$ are given by
\begin{align}
    \frac{d\tilde{G}^{(0)}_m}{dA} = \sum_{n=m}^{\infty} \Big(\frac{BA^2}{\rho^2} \Big)^n & f_{nm} \Bigg[  \Big(\frac{2n}{A}-\frac{2nA}{\rho^2} \Big) {}_2F_1(\frac{1}{2}-n,-n; \frac{1}{2}-m-n; \frac{\rho^2}{A^2}) \nonumber\\ 
    + & \Big(-\frac{2\rho^2}{A^3} + \frac{2}{A} \Big) \frac{(\frac{1}{2}-n)(-n)}{(\frac{1}{2}-m-n)} {}_2F_1(\frac{3}{2}-n,1-n; \frac{3}{2}-m-n; \frac{\rho^2}{A^2})  \Bigg]
\end{align}
\begin{align}
    \frac{d^2\tilde{G}^{(0)}_m}{dA^2} = \sum_{n=m}^{\infty} \Big(\frac{BA^2}{\rho^2} \Big)^n & f_{nm} \Big( \frac{\rho^2-A^2}{\rho^4} \Big) \Bigg[(2m-1-4n(m+1))\  {}_2F_1(\frac{1}{2}-n,-n;\frac{1}{2}-m-n;\frac{\rho^2}{A^2}) \nonumber\\
    &+ (1-4n^2+m(4n-2))\ {}_2F_1(\frac{3}{2}-n,-n;\frac{1}{2}-m-n;\frac{\rho^2}{A^2})\Bigg].
\end{align}
These are sufficient to calculate all terms up to quadratic order in the bias expansion. In our fiducial cosmology at $z = 0$ and along the line of sight ($\nu = 1$) we find that the sums in $\tilde{G}_m$ converge at the sub-percent level within thirty summands.

From the above, contributions to the redshift-space Zeldovich power spectrum can be calculated using spherical Bessel transforms of the specific form
\begin{equation}
    P_s(\bk) \ni 4\pi \sum_{\ell=0}^{\infty}\ \int dq\ q^2\ e^{-\frac{1}{2}K^2(X+Y)}\ \Big(\frac{-2}{kq} \Big)^\ell \ \tilde{G}^{(n)}_\ell(kqc,-\frac{1}{2}K^2Y,kq)\ \mathcal{A}_n(q) \ j_\ell(kq),
\end{equation}
where the scalar function $\mathcal{A}_n$ are tabulated in Table~\ref{table:MII_table}.

\begin{table}[h!]
    \centering
        \begin{tabular}{c|c|c|c|c}
        $\mathcal{A}_n$ & $n = 0$ & $n = 1$ & $n = 2$ & ... \\
        \hline
        $1$ & $1$ & 0 & 0 &  \\
        $b_1$ & 0 & $2 K U(q)$ & 0 &  \\
        $b_1^2$ & $\xi_L(q)$ & 0 & $K^2 U(q)^2$ & \\
        $b_2$ & 0 & 0 & $K^2 U(q)^2$ & \\
        $b_1 b_2$ & 0 & $K U(q) \xi_L(q)$ & 0 & \\
        $b_2^2$ & $\frac{1}{2} \xi_L(q)^2$ & 0 & 0 & \\
        $b_s$ & $-K^2 X_{s^2}(q)$ & 0 & $K^2 Y_{s^2}(q)$ & \\
        $b_1 b_s$ & 0 & $2K V(q)$ & 0 & \\
        $b_2 b_s$ & $\chi^{12}(q)$ & 0 & 0 & \\
        $b_s^2$ & $\zeta(q)$ & 0 & 0 & 
    \end{tabular}

    \caption{Table of power spectrum contributions in M{\sc ii}.}
    \label{table:MII_table}
\end{table}

\section{Wiggle/No-Wiggle split}
\label{app:wiggles}

Most analyses of BAO data to date have employed empirical models for the post-reconstruction power spectrum or correlation function often motivated by theoretical calculations and calibrated to N-body simulations. Refs.~\cite{Vlah16,Ding18} showed that the analytic form of these empirical models can be interpreted within perturbation theory as a resummation of bulk displacements at the BAO scale. In this appendix we re-derive their results within our Zeldovich calculation, updating the scale dependences and redshift-space factors where appropriate.

Let us first examine the displaced-displaced cross spectrum in redshift space. Following refs.~\cite{Vlah16,Ding18} we split the displacement two-point function into $A^{dd}_{ij} = A^{dd, \rm nw}_{ij} + \Delta A^{dd,\rm w}_{ij}$, where the no-wiggle and wiggle pieces are calculated by substituting $P_{\rm nw}$ and $\Delta P_{\rm w}$ into Equation~\ref{eqn:XYdef}. Making the assumption that the latter, $\Delta A^{dd,\rm w}_{ij}$, is small enough as to be perturbative\footnote{Taking the nonlinear scale to be given by $k_{\rm nl}^2 \Sigma^2(z) \sim 1$, we have $k_{\rm nl}\propto D^{-1}(z)$, such that the peak magnitude of $k_i k_j \Delta A^{\rm w}_{ij}$ is roughly in the few tenths of a percent range for our reference cosmology independent of redshift.}, we can Taylor expand the exponential in the Zeldovich integrand to get
\begin{equation}
    P^{dd}(\bk) = \dtq e^{-i \bk \cdot \bq - \frac{1}{2} K_i K_j A^{dd, \rm nw}_{ij} }\ \Big(1 - \frac{1}{2} K_i K_j \Delta A^{dd, \rm w}_{ij} + \mathcal{O}(k^4 \Sigma^4) \Big) \Big(1 + 2 i b_1 K_i U^{d}_i(q) + ... \Big), \nonumber
\end{equation}
where we have used the transformed $K_i = R_{ij} k_j$ to encode redshift-space effects. Given that the no-wiggle spectrum reproduces the broadband scale dependence of the linear theory power spectrum, we can think of the no-wiggle exponential as resumming the non-BAO component of large scale bulk flows.  Since the wiggle component contributes negligibly to the displacement power in the perturbative limit, keeping only one power of the wiggle power spectrum in our calculations serves to distinguish the effect of the IR bulk flows from BAO phenomena. The two-point functions entering the bias terms can likewise be split into no-wiggle and wiggle pieces, e.g.\ $U(q) = U^{\rm nw} + \Delta U^{\rm w}$, where again, roughly speaking, the former will contribute only to to the broadband power while the latter will give rise to oscillatory behavior. Keeping the above expression to order\footnote{Note, however, that terms involving more powers of the wiggle displacement will be more suppressed than those involving no-wiggle displacements.} $\mathcal{O}(k^2 \Sigma^2)$, and discarding terms that don't contain any no-wiggle pieces, we then have
\begin{align}
    P^{dd}(\bk) &\ni \dtq e^{-i \bk \cdot \bq - \frac{1}{2} K_i K_j A^{dd, \rm nw}_{ij} }\ \Big(- \frac{1}{2} K_i K_j \Delta A^{dd, \rm w}_{ij} + 2 i b_1 K_i \Delta U^{d,\rm w}_i(q) + b_1^2 \xi^{\rm w}_L(q) + ... \Big) \nonumber \\
    &\approx  e^{-\frac{1}{2} K_i K_j  \bar{A}^{dd, \rm nw}_{ij} } \dtq e^{-i \bk \cdot \bq }\ \Big(- \frac{1}{2} K_i K_j \Delta A^{dd, \rm w}_{ij} + 2 i b_1 K_i \Delta U^{d,\rm w}_i(q) + b_1^2 \xi^{\rm w}_L(q) + ... \Big) \nonumber
\end{align}
where in the final line we have used the fact that the wiggle contributions will be confined in support around the BAO scale ($q \sim 100\,$Mpc) and the non-wiggle pieces vary smoothly at this Lagrangian separation, so we can pull the exponentiated no-wiggle contribution out of the integral as an average.  Following ref.~\cite{Vlah16}, we have defined the quantity $\bar{A}^{dd, \rm nw}_{ij}$ as the ``average'' of the un-barred quantity over the support of the wiggle component; to zeroth order in the approximation this is equivalent to evaluating $A_{ij}$ at the peak $q_{\rm max}$ of the support of the wiggle feature. Neglecting any angular effects in $\nu = \hq \cdot \hat{k}$, which will enter at higher order in the wave number, we further have that $\bar{A}^{dd, \rm nw}_{ij} \simeq (X^{dd, \rm nw} + \frac{1}{3}Y^{dd, \rm nw})\delta_{ij}$\footnote{\label{fn:damp_fac}The factor of a third, included also in ref.~\cite{Vlah16} but not in ref.~\cite{Ding18}, comes from the angular average $\langle \hq_i \hq_j \rangle = \delta_{ij}/3$. This can be justified by noting that the integral
\begin{equation}
    \frac{1}{2} \int\ d\mu\ e^{ikq\mu - \frac{1}{2}k^2 \mu^2 Y/2} = e^{-k^2 Y/6} j_0(kq) + \mathcal{O}(k^4\Sigma^4)
\end{equation}
We note, however, that this prescription is only approximate; for example, the same integral with an additional factor of $\mu$ in the integrand, relevant for the $b_1$ contribution, would instead yield $\exp[-3k^2Y/10]\,j_1(kq)$ at leading order. In general, bias contributions with more angular dependence will be damped more.  This effect is automatically included in the full Zeldovich calculation.}.
Plugging in for the expression of $\textbf{K} = \textbf{R}^{\textbf{T}} \bk$, with $K^2 = (1 + f(f+2)\nu^2)k^2$, the wiggle contribution to the power spectrum is then approximately
\begin{align}
    P^{dd}(\bk)_{\rm wiggle} &\approx e^{-\frac{1}{2} K^2 \Sigma_{dd}^2} \left[ \big(\hat{K} \cdot \hat{k}\big)^2  P^{dd, \rm w}(k) + 2 b_1 \big(\hat{K} \cdot \hat{k}\big)  P^{dm, \rm w}(k) + b_1^2   P^{mm, \rm w}(k) \right] \nonumber \\
    &= e^{-\frac{1}{2} K^2 \Sigma_{dd}^2} \left[ (1 + f\nu^2)^2 (1 - \mathcal{S}(k))^2 + 2 b_1 (1 + f\nu^2)(1 - \mathcal{S}(k)) + b_1^2 \right] P_{\rm w}(k) \nonumber \\
    &= e^{-\frac{1}{2} K^2 \Sigma_{dd}^2} \left[ (1+f\nu^2)(1-\mathcal{S}(k)) + b_1 \right]^2 P_{\rm w}(k)
    \label{eqn:pdd_approx}
\end{align}
where in the penultimate equality we have used the definition of the displaced field and defined $\Sigma_{dd}^2 = (X^{dd, \rm nw} + \frac{1}{3} Y^{dd, \rm nw})(q_{max})$ to be evaluated at the peak of the wiggle displacements. This recovers the form of the empirical model in ref.~\cite{Ding18} when we take the Eulerian bias to be $b_1^E = 1 + b_1$, and stick to the damping expansion approximation introduced in ref.~\cite{Vlah16}. Explicit expressions for $X$ and $Y$ are given in Equation~\ref{eqn:Xds}. Taking $S \rightarrow 0$ in the above expression gives the unreconstructed power spectrum within this approximation.

We can now derive the analytical form of the reconstructed power spectrum for \textbf{Rec-Sym} in this approximation. Explicitly, we have
\begin{align}
    P^{ds}_{\rm wiggle}(\bk) &= -e^{-\frac{1}{2} K^2 \Sigma_{ds}^2} \Big( (1+f\nu^2)(1-\mathcal{S}(k)) + b_1 \Big) (1+f\nu^2) \mathcal{S}(k)  P_{\rm w}(k) \\
    P^{ss}_{\rm wiggle}(\bk) &= e^{-\frac{1}{2} K^2 \Sigma_{ds}^2} (1 + f\nu^2)^2 \mathcal{S}(k)^2 P_{\rm w}(k).
\end{align}
The $\Sigma^2_{ab}$ are defined as in the $dd$ case. These are the same expressions as derived in ref.~\cite{Ding18}, though we differ on the expressions for the $\Sigma_{ab}^2$ that are involved. Our expressions also agree with those in refs.~\cite{Coh16,Sherwin19} in the limit that $q_{max} \rightarrow \infty$, though we note that this limit doesn't as accurately capture the damping of the feature since it resums the IR displacements at $q$ beyond the BAO scale. Adding the three spectra together, we recover the Kaiser limit as $k \rightarrow 0$, with different damping factors entering at different scales via $\Sigma^2_{ab}$. Note that in \textbf{Rec-Sym} the angular dependence of the damping is identical in each piece and is encoded within the $\nu$ dependence of $K^2$.

The reconstructed power spectrum with $\textbf{Rec-Iso}$ requires a few additional modifications. The displaced-displaced auto spectrum is unchanged, and the shifted-shifted auto spectrum can be calculated by setting $f = 0$ in all formulae, as noted in the main body of the text. However, the $ds$ cross spectrum requires more care, since the zero lag pieces do not transform equally. In particular, direct inspection of the exact expression in Equation~\ref{eqn:methodii_cross} shows that we should instead define
\begin{equation}
    - \frac{1}{2} k^2 \Sigma^2_{ds, \rm iso} = - \frac{1}{2} k^2 \left[ (1 + f(f+2)\nu^2) \Sigma^{(dd)} + \Sigma^{(ss)} - 2 (1 + f\nu^2)\ \Big( \tilde{X}^{ds} + \frac{1}{3} \tilde{Y}^{ds} \Big) \right]_{q=q_{\rm max}} \quad .
\end{equation}
Note this expression differs in detail from that in ref.~\cite{Ding18}.  The cross spectrum is then instead
\begin{equation}
    P^{ds, \rm iso}_{\rm wiggle}(\bk) = - e^{-\frac{1}{2} k^2 \Sigma_{ds,{\rm iso}}^2} \left[ (1+f\nu^2)(1-\mathcal{S}(k)) + b_1 \right] \mathcal{S}(k)  P_{\rm w}(k),
\end{equation}
where the angular dependence is subsumed into the defintion of $\Sigma_{ds, \rm iso}$. Unlike \textbf{Rec-Sym}, the damping factor in \textbf{Rec-Iso} is not captured by a single angular dependence.

We end this section with a discussion of the inclusion of higher bias terms and other corrections. As seen in the main body of the text, higher bias terms $b_2$ and $b_s$, incorporated in our Zeldovich calculation, contribute not only to the broadband but also serve to shift and smear the BAO feature itself. It might thus be of interest to extend the above approximation to include also these higher bias contributions. A potential avenue has been highlighted in ref.~\cite{Ding18}, although an approach closer to our perturbative bias expansion could also be explored.

Finally, the calculation in ref.~\cite{Ding18} included a derivative bias, $b_{\nabla^2} \nabla^2 \delta$, as a proxy to estimate the contributions of the higher bias operators. These derivative bias terms can easily be included in the above expressions by substituting $b_1 \rightarrow b_1 + k^2 b_{\nabla^2}$. However, there is another context in which such a term might arise in which it would differ across the three pieces $dd$, $ds$ and $ss$: if the smoothing due to the $\Sigma^2_{ab}$'s as defined above do not accurately capture the IR bulk flows -- for example if the broadband properities of $P_{\rm nw}$ are slightly off -- but differ by some perturbatively small $k^2 \delta \Sigma^2_{ab}$, the resulting correction could be corrected for by terms of the form $c_{ab}^2 k^2 P^{ab, \rm w}(k)$, where $c_{ab}^2$ would constants fit individually to $dd$, $ds$ and $ss$. Such corrections are essentially identical to the EFT corrections described in the text for the full Zeldovich calculation.

\section{Nonlinearities from the Lagrangian to Eulerian mapping}
\label{app:EulerianPosition}

In standard density field reconstruction, each galaxy is shifted by a smoothed displacement field $\chi$ evaluated at the galaxy's current \textit{Eulerian} position $\bx = \bq + \bPsi(\bq)$ (Equation~\ref{eqn:deltak_ds}). In the main body of the text, we worked in the approximation that $\chi(\bx) \approx \chi(\bq)$, with the understanding that nonlinear corrections would be suppressed by the smoothing scale $\sim \bPsi/R$. The goal of this appendix is to flesh out this statement by explicitly computing the leading order corrections to the reconstructed matter power spectrum due to the mapping nonlinearity in real space. For the sake of brevity we will defer the effects of other nonlinearities, such as those arising from dynamics or from translating between displacements and densities, to future work. Earlier treatments of such effects in Eulerian perturbation theory can be found in refs.~\cite{Sch15,Hik17}.

Assuming that the shift field $\chi(\bq)$ defined in Lagrangian space is Gaussian, the displaced field with mapping nonlinearities unsuppressed is given by
\begin{equation}
    \tilde{\bPsi}^d_i(\bq) = \bPsi^d_i(\bq) + \bPsi_n \partial_n \chi_i(\bq) + \frac{1}{2} \bPsi_n \bPsi_m \partial_n \partial_m \chi_i(\bq) + ... \equiv \bPsi^d_i + \bPsi^{(d,2)}_i + \bPsi^{(d,3)}_i + ...
\end{equation}
where we have kept the convention used in the main text to refer to the linear piece as $\bPsi^d = (1 - \mathcal{S}) \ast \bPsi$, referring to the nonlinear field as $\tilde{\bPsi}^d = \bPsi(\bq) + \chi(\bq + \bPsi)$. For the remainder of this appendix we will focus on corrections due to $\bPsi^{(d,2)}$.\footnote{At one loop order all corrections due to $\bPsi^{(d,3)}$ are degenerate with the counterterms in our model. To see this, note that such corrections contractions with linear displacements, e.g.
\begin{equation}
    \langle \bPsi^{a}_{1,i} \bPsi^{(d,3)}_{2,j} \rangle  = \langle \bPsi_m \partial_m \partial_n \chi_j \rangle \langle \bPsi^a_{1,i} \bPsi_{2,n} \rangle + \frac{1}{2} \langle \bPsi_n \bPsi_m \rangle \langle \bPsi^a_{1,i} \partial_n \partial_m \chi_{2,j} \rangle, \; a = d, s.
\end{equation}
Multiplied by the appropriate factor of $-\frac{1}{2} k_i k_j$, the two pieces on the right hand side Fourier transform into $\sim k^2 P^{am}_L(k)$ and $\sim k^2 P^{as}_L(k)$, respectively, thus taking the form of our counterterms $\sim k^2 P^{ab}(k)$.
}

From the above,we can write the nonlinear displaced-displaced autospectrum as
\begin{equation}
    P^{dd}(k) - P^{dd}_{\rm Zel}(k) = \dtq e^{i \bk \cdot \bq - \frac{1}{2} k_i k_j A^{dd}_{ij}} \Big( \exp\Big[-\frac{1}{2}k_i k_j A^{dd,\rm 1-loop}_{ij} - \frac{i}{6} k_i k_j k_k W^{dd}_{ijk} \Big] - 1 \Big) + \mathcal{O}(P_L^3) \nonumber 
\end{equation}
where we have defined
\begin{align}
    &A^{dd,\rm 1-loop}_{ij} = \avg{ \Delta^{(dd,2)}_i \Delta^{(dd,2)}_j  }_c + 2\avg{ \Delta^{(dd,1)}_i \Delta^{(dd,3)}_j  }_c\nonumber \\
    &W^{dd}_{ijk} = \avg{ \Delta^{(dd,1)}_i \Delta^{(dd,1)}_j \Delta^{(dd,2)}_k  }_c + (121) + (211).
\end{align}
as in the case of the nonlinear matter power spectrum (e.g. \cite{VlaWhiAvi15}). To calculate these we note that
\begin{align}
    \langle \bPsi^{(d,2)}_i(\bq_2) &\bPsi^{(d,2)}_j(\bq_1) \rangle_c = \avg{\big(\bPsi_n \partial_n \chi_i \big)(\bq_2) \big(\bPsi_m \partial_m \chi_j \big)(\bq_1)  }_c \nonumber \\
    &= \avg{\bPsi_n(\bq_2) \bPsi_m(\bq_1)} \avg{\partial_n \chi_i(\bq_2) \partial_m \chi_j(\bq_1)} + \avg{\bPsi_n(\bq_2) \partial_m \chi_j(\bq_1)} \avg{\bPsi_m(\bq_1) \partial_n \chi_i(\bq_2)} \nonumber 
\end{align}
and
\begin{align}
    W^{dd, 112}_{ijk} =2 \avg{\bPsi^d_{1,i} \bPsi_{2,n} - \bPsi^d_{i} \bPsi_{n}} \avg{\bPsi^{d}_{1,j} \partial_n \chi_{2,k}} +  \Big( i \leftrightarrow j \Big)
    \label{eqn:wdd}
\end{align}
where numerical subscripts refer to coordinates $\bq_{1,2}$ as usual.

The mapping corrections to the cross spectrum $P^{ds}$ can be similarly calculated. In this case we need the displacement correlators
\begin{align}
    A^{ds,22}_{ij} = \avg{\bPsi^{(d,2)}_i \bPsi^{(d,2)}_j} = \avg{\bPsi_n \bPsi_m} \avg{\partial_n \chi_i \partial_m \chi_j}
    \label{eqn:Ads22}
\end{align}
where all expectation values are evaluated at a point since $\bPsi^s$ receives no nonlinear corrections from the Eulerian-Lagrangian mapping and similarly
\begin{align}
    W^{ds,112}_{ijk} &= \avg{ \bPsi^s_{i,1} \bPsi^s_{j,1} \bPsi^{(d,2)}_{k,2} } - \avg{\bPsi^d_{i,2} \bPsi^s_{j,1} \bPsi^{(d,2)}_{k,2}} - \avg{\bPsi^s_{i,1} \bPsi^d_{j,2} \bPsi^{(d,2)}_{k,2}} \nonumber \\
    &= \Big( \avg{\bPsi^s_{1,i} \bPsi_{2,n}} - \avg{\bPsi^d_{i} \bPsi_{n}} \Big) \avg{\bPsi^s_{1,j} \partial_n \chi_{2,k}} + \Big( i \leftrightarrow j \Big)
\end{align}
to one loop order.

\begin{figure}
    \centering
    \includegraphics[width=\textwidth]{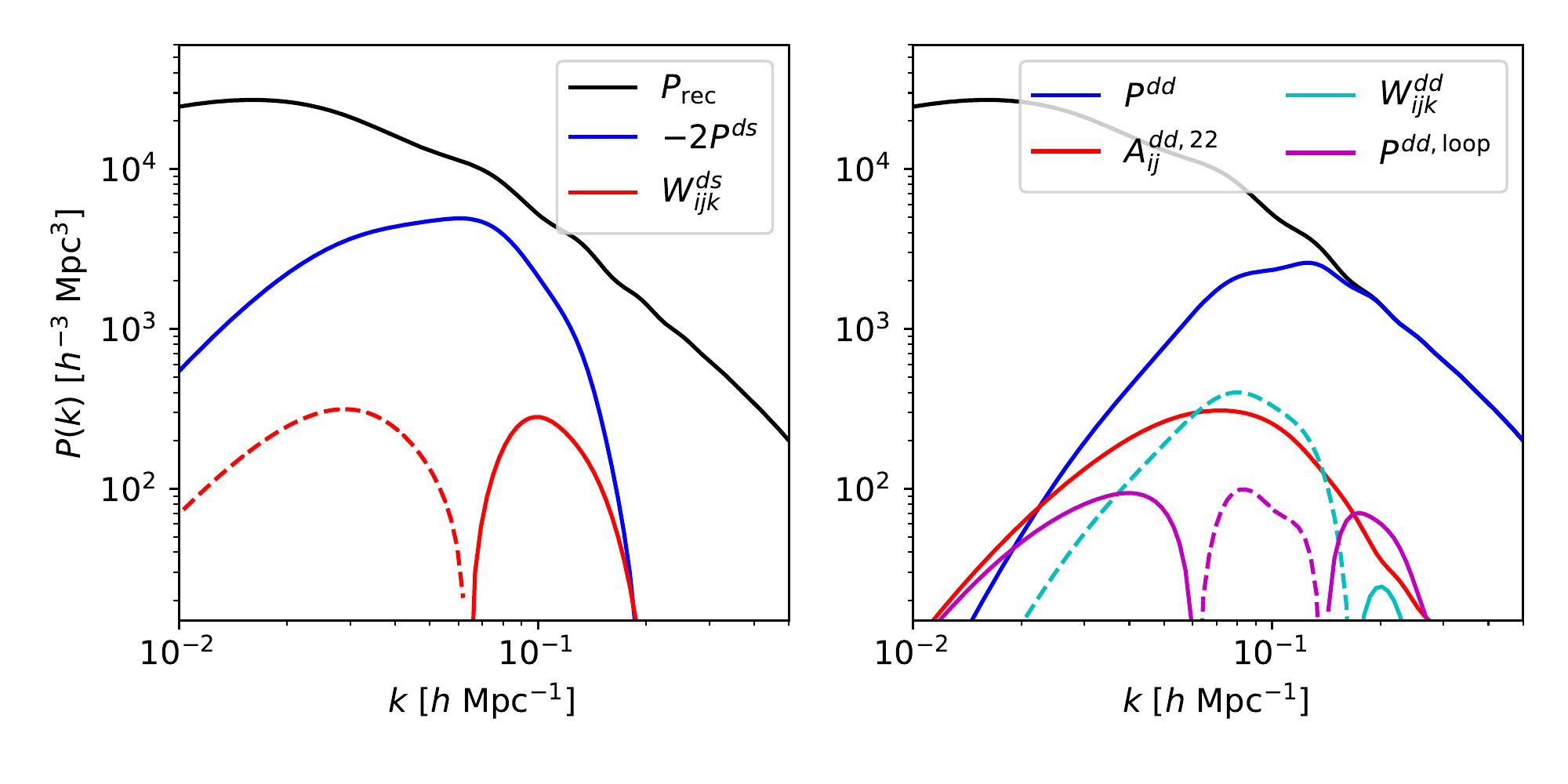}
    \caption{Nonlinear corrections to the reconstructed matter power spectrum due to the Lagrangian-to-Eulerian mapping at one loop order, for $z=0$ and $R = 15\ h^{-1}$ Mpc. The left and right panels show contributions to the $ds$ and $dd$ power spectra, respectively. Even for the worst case of $z=0$, the corrections are never more than a few percent of the total reconstructed power spectrum, though they can become larger than the constituent $dd, ds$ spectra at large or small scales.}
    \label{fig:mapping_corrs}
\end{figure}

As might have been expected, the mapping corrections above all take the form of products of displacement two point functions and their derivatives. Roughly speaking these corrections each have amplitudes given by powers of the Zeldovich displacement $\Sigma^2$ and wavenumber $k$ capped at $R^{-1}$ by the smoothing filter; we can thus expect the corrections to enter at order $(\Sigma/R)^4$. For a $15 h$ Mpc$^{-1}$ filter at $z = 0$ this amounts to a percent-level effect, with smaller effects at higher $z$. Concretely, these two point functions can be calculated using
\begin{align}
    &\avg{\partial_n \chi_i(\bq_2) \partial_m \chi_j(\bq_1)} = \mathcal{A}^{ss}(q) (\delta_{ij} \delta_{nm} + ...\ ) + \mathcal{B}^{ss}(q) (\hq_i \hq_j \delta_{nm} + ...\ ) + \mathcal{C}^{ss}(q) \hq_i \hq_j \hq_n \hq_m \nonumber \\
    &\avg{\bPsi_n(\bq_2) \partial_m \chi_i(\bq_1)} = \mathcal{D}^{sm}(q) (\hq_i \delta_{nm} + ...\ ) + \mathcal{E}^{sm}(q) \hq_i \hq_n \hq_m \nonumber \\
    &\avg{\bPsi^a_n(\bq_2) \partial_m \chi_i(\bq_1)} = \mathcal{D}^{as}(q) (\hq_i \delta_{nm} + ...\ ) + \mathcal{E}^{as}(q) \hq_i \hq_n \hq_m \nonumber
\end{align}
where the ellipses denote all distinct permutations and the scalar functions are given by
\begin{align}
    &\mathcal{A}^{ss}(q) = \frac{1}{105} \int \frac{dk\ k^2}{2\pi^2}\ \big(7 j_0(kq) + 10 j_2(kq) + 3 j_4(kq) \big) P^{ss}_L(k) \nonumber \\
    &\mathcal{B}^{ss}(q) = - \frac{1}{7} \int \frac{dk\ k^2}{2\pi^2}\ \big(j_2(kq) + j_4(kq) \big) P^{ss}_L(k) \nonumber \\
    &\mathcal{C}^{ss}(q) = \int \frac{dk\ k^2}{2\pi^2} j_4(kq)  P^{ss}_L(k) \nonumber \\
    &\mathcal{D}^{ab}(q) = \frac{1}{5} \int \frac{dk\ k}{2\pi^2}\ \big(j_1(kq) + j_3(kq) \big) P^{ab}_L(k) \nonumber \\
    &\mathcal{E}^{ab}(q) = - \int \frac{dk\ k}{2\pi^2}\ j_3(kq) P^{ab}_L(k) 
    \label{eqn:scalar_comps}
\end{align}
where we have used the identification $\chi = \bPsi^s.$ The remaining correlator $\avg{\bPsi_{2,i} \bPsi_{1,j}}$ is simply minus the non-zero lag piece of $A_{ij}$. Finally, when some or all of the displacement correlators in each product are contracted at the same point, as for example in the first term in Equation~\ref{eqn:wdd} and Equation~\ref{eqn:Ads22}, the resulting contribution becomes proportional to $k^2 P_{ab, \rm Zel}$ and degenerate with the counterterms included in our model. The above corrections from the nonlinear Eulerian-Lagrangian mapping are plotted at $z = 0$ for the smoothing scale $R = 15 h^{-1}$ Mpc in Figure~\ref{fig:mapping_corrs}. As expected, even at $z =0$ they are never more than a few percent of the total reconstructed power, though interestingly they can become comparable or larger than the Zeldovich $P^{dd}$ and $P^{ds}$ individually on scales where the Zeldovich spectra lose support. We caution that these curves do not include comparable corrections due to bias or dynamical nonlinearities.

We close with some general comments about nonlinearities in reconstruction. Firstly, the mapping corrections enumerated above are not the only ones at one-loop order; by focusing only on corrections due to $\bPsi^{(d,2)}$ we have explicitly avoided the $(13)$ contributions due to third-order mapping corrections. Moreover, as this was an exploratory exercise with which to evaluate the magnitude of mapping nonlinearities, we chose not to include the effects of bias, which would require the inclusion of terms such as $\langle \delta_1^2 \bPsi^{(d,2)}_i \rangle$, though these will be in general decomposable into components much like those in Equation~\ref{eqn:scalar_comps}. Finally, in addition to mapping nonlinearities, by approximating the shift vector $\chi$ with the smoothed Zeldovich displacement we have ignored nonlinearities induced by translating between the density field and displacements. We expect these will be of similar importance to the mapping corrections but defer their evaluation for future work, noting that only that both nonlinearities can be trivially reduced by pushing the smoothing scale $R$ deeper into the linear regime. That these effects are expansions in $\Sigma/R$ distinguishes them from nonlinear bias or dynamics.

\bibliographystyle{JHEP}
\bibliography{main}
\end{document}